%% file: icse2026-p1626.tex
\PassOptionsToPackage{table,xcdraw}{xcolor}
\documentclass[sigconf,screen]{acmart}
\AtBeginDocument{%
  }

\usepackage{amsmath,amsfonts}
\usepackage{algorithmic}
\usepackage{xcolor}

\usepackage{threeparttable} 
\usepackage{enumitem}
\usepackage{array}
\usepackage{wrapfig}
\usepackage{amsmath,amsfonts}
\usepackage{graphicx}
\usepackage{textcomp}
\usepackage{float}
\usepackage{listings}
\usepackage{xspace}
\usepackage{multirow}
\usepackage{amsthm}

\usepackage{balance}
\usepackage{algorithm}
\usepackage{colortbl}
\usepackage{subcaption}

\usepackage[skins]{tcolorbox}
\usepackage{xcolor}
\usepackage{multicol}
\usepackage{soul}
\usepackage{framed}
\usepackage{booktabs}
\usepackage{booktabs}
\usepackage{multirow}
\usepackage{siunitx}
\newcommand*\colourcheck[1]{%
	\expandafter\newcommand\csname #1check\endcsname{\textcolor{#1}{\ding{52}}}%
}
\colourcheck{blue}
\colourcheck{green}
\colourcheck{red}
\newtcolorbox{boxB}[2][]{%
  enhanced,colback=white,colframe=black,coltitle=black,
  sharp corners,
  toprule=1.0pt,
  rightrule=0.3pt,
  leftrule=0pt,
  bottomrule=0pt,
  fonttitle=\itshape\scshape\large,
  left=0pt,right=5pt,top=5pt,bottom=3pt,
  attach boxed title to top right={yshift=-0.3\baselineskip-0.4pt,xshift=-5mm},
  boxed title style={tile,size=minimal,left=0.2mm,right=0.5mm,
    colback=white,before upper=\strut},
  title=#2,#1
}

\definecolor{sh_string}{rgb}{0.1, 0.6, 0.1}

\lstdefinelanguage{json}{
  basicstyle=\ttfamily,
  showstringspaces=false,
  morestring=[b]",
  stringstyle=\color{sh_string}
}

\newcommand{\tool}{\textsc{RBCTest}\xspace}

\newboolean{showcomments}
\setboolean{showcomments}{true}
\ifthenelse{\boolean{showcomments}}
 { \newcommand{\mynote}[2]{
      \fbox{\bfseries\sffamily\scriptsize#1}
        {\small$\blacktriangleright$\textsf{\emph{#2}}$\blacktriangleleft$}}}
        { \newcommand{\mynote}[2]{}}

\usepackage{tikz}

\newcommand{\circlednum}[2][red]{%
\tikz[baseline=(char.base)]{
    \node[shape=circle, fill=#1, text=white, inner sep=1pt] (char) {#2};}%
}

\definecolor{request}{HTML}{D80073}
\definecolor{response}{HTML}{0050EF}

\newcolumntype{L}[1]{>{\raggedright\arraybackslash}p{#1}}
\newtheorem{Observation}{Observation}

\newcommand{\code}[1]{{\footnotesize\texttt{#1}}}
\usepackage{amsthm}
\definecolor{dkgreen}{rgb}{0,0.6,0}
\definecolor{gray}{rgb}{0.5,0.5,0.5}
\definecolor{lightgray}{rgb}{211, 211, 211}
\definecolor{mauve}{rgb}{0.58,0,0.82}

\definecolor{custom-red}{rgb}{0,0,0}
\definecolor{custom-blue}{rgb}{0,0,0}

\definecolor{c1}{HTML}{f4cccc}
\definecolor{c2}{HTML}{f5cdcd}
\definecolor{c3}{HTML}{fffcfc}
\definecolor{c4}{HTML}{ffffff}
\definecolor{c5}{HTML}{ffffff}
\definecolor{c6}{HTML}{fffdfd}
\definecolor{c7}{HTML}{f5cfcf}
\definecolor{c8}{HTML}{fffbfb}
\definecolor{c9}{HTML}{ffffff}
\definecolor{c10}{HTML}{fffdfd}
\definecolor{c11}{HTML}{fefafa}
\definecolor{c12}{HTML}{fef7f7}
\definecolor{c13}{HTML}{ffffff}
\definecolor{c14}{HTML}{fffefe}
\definecolor{c15}{HTML}{ffffff}
\definecolor{c16}{HTML}{fefafa}
\definecolor{c17}{HTML}{fdf3f3}
\definecolor{c18}{HTML}{fffefe}
\definecolor{c19}{HTML}{fdf5f5}
\definecolor{c20}{HTML}{ffffff}

\definecolor{codegreen}{rgb}{0,0.6,0}
\definecolor{codegray}{rgb}{0.5,0.5,0.5}
\definecolor{codepurple}{rgb}{0.58,0,0.82}
\definecolor{backcolour}{rgb}{0.95,0.95,0.92}

\lstdefinestyle{jsonStyle}{
    backgroundcolor=\color{white},
    commentstyle=\color{codegreen},
    keywordstyle=\color{blue},
    numberstyle=\tiny\color{codegray},
    stringstyle=\color{codepurple},
    basicstyle=\footnotesize\ttfamily,
    breakatwhitespace=false,
    breaklines=true,
    captionpos=b,
    keepspaces=true,
    numbers=left,
    numbersep=5pt,
    showspaces=false,
    showstringspaces=false,
    showtabs=false,
    tabsize=2,
    language=JSON,
}
\definecolor{lightblue}{RGB}{204, 212, 244}

\definecolor{json_key}{RGB}{0,0,255}         
\definecolor{json_string}{RGB}{160,32,240}   
\definecolor{json_number}{RGB}{0,128,0}      
\definecolor{json_boolean}{RGB}{255,0,0}     

\lstdefinestyle{jsonStyle}{
    backgroundcolor=\color{white},
    commentstyle=\color{gray},
    keywordstyle=\color{json_key},
    numberstyle=\tiny\color{gray},
    stringstyle=\color{json_string},
    basicstyle=\footnotesize\ttfamily,
    breakatwhitespace=false,
    breaklines=true,
    captionpos=b,
    keepspaces=true,
    numbers=left,
    numbersep=5pt,
    showspaces=false,
    showstringspaces=false,
    showtabs=false,
    tabsize=2,
    language=json,
    literate=
     *{:}{{\textcolor{json_key}{:}}}1
      {,}{{\textcolor{gray}{,}}}1
      {"}{{\ProcessQuote{"}}}1
      {'}{{\ProcessQuote{'}}}1
      {true}{{\textcolor{json_boolean}{true}}}4
      {false}{{\textcolor{json_boolean}{false}}}5
      {null}{{\textcolor{json_boolean}{null}}}4
      {\{}{{\textcolor{gray}{\{}}}1
      {\}}{{\textcolor{gray}{\}}}}1
}

\newcommand*{\ProcessQuote}[1]{\textcolor{json_string}{#1}}

\lstdefinestyle{yaml}{
     basicstyle=\color{blue}\footnotesize,
     rulecolor=\color{black},
     string=[s]{'}{'},
     stringstyle=\color{blue},
     comment=[l]{:},
     commentstyle=\color{black},
     morecomment=[l]{-}
 }

\definecolor{codegray}{gray}{0.5}
\definecolor{codepurple}{rgb}{0.58,0,0.82}
\definecolor{backcolour}{rgb}{0.95,0.95,0.92}
\definecolor{blue}{rgb}{0,0,1}

\definecolor{braceblue}{rgb}{0,0,0.8}

\lstdefinestyle{mystyle}{
    basicstyle=\ttfamily,
    columns=fullflexible,
    keepspaces=true,
    showstringspaces=false,
    commentstyle=\color{braceblue},
    keywordstyle=\color{braceblue},
    stringstyle=\color{braceblue},
    moredelim=**[s][\color{braceblue}]{{}{}},
}

\lstdefinestyle{pythonStyle}{
    backgroundcolor=\color{backcolour},
    commentstyle=\color{codegray},
    keywordstyle=\color{magenta},
    numberstyle=\tiny\color{codegray},
    stringstyle=\color{codepurple},
    basicstyle=\ttfamily\footnotesize,
    breakatwhitespace=false,
    breaklines=true,
    captionpos=b,
    keepspaces=true,
    numbers=left,
    numbersep=5pt,
    showspaces=false,
    showstringspaces=false,
    showtabs=false,
    tabsize=2,
    language=Python,
}
\newcolumntype{?}{!{\vrule width 0.5pt \hspace{0.05pt} \vrule width 0.5pt}}

\copyrightyear{2026}
\acmYear{2026}
\setcopyright{cc}
\setcctype{by-nc-nd}
\acmConference[ICSE '26]{2026 IEEE/ACM 48th International Conference on Software Engineering}{April 12--18, 2026}{Rio de Janeiro, Brazil}
\acmBooktitle{2026 IEEE/ACM 48th International Conference on Software Engineering (ICSE '26), April 12--18, 2026, Rio de Janeiro, Brazil}
\acmPrice{}
\acmDOI{10.1145/3744916.3787775}
\acmISBN{979-8-4007-2025-3/2026/04}

\begin{document}





\title {{\tool}: Leveraging LLMs to Mine and Verify Oracles of\\ API Response Bodies for RESTful API Testing}




\author{Hieu Huynh}
\authornote{Co-first authors, equal contributions}
\orcid{0000-0003-2310-1376}
\affiliation{%
 \institution{Katalon Inc.}
 \city{Ho Chi Minh city}
 \country{Vietnam}
}
\email{minhhieu2214@gmail.com}

\author {Quoc-Tri Le}
\authornotemark[1]
\orcid{0009-0006-2335-7282}
\affiliation{%
 \institution{Katalon Inc.}
 \city{Ho Chi Minh city}
 \country{Vietnam}
}
\email{lqtri691@gmail.com}

\author {Tu Nguyen}
\authornotemark[1]
\orcid{0009-0000-7307-0171}
\affiliation{%
 \institution{University of Science}
 \institution{Vietnam National Univ.}
 \city{Ho Chi Minh city}
 \country{Vietnam}
}
\email{23C15017@student.hcmus.edu.vn}

\author {Viet Nguyen}
\orcid{0009-0000-7579-8991}
\affiliation{%
 \institution{University of Science}
 \institution{Vietnam National Univ.}
 \city{Ho Chi Minh city}
 \country{Vietnam}
}
\email{21C11045@student.hcmus.edu.vn}

\author{Vu Nguyen}
\authornote{Corresponding author.}
\affiliation{%
 \institution{University of Science}
 \institution{Vietnam National Univ.}
 \city{Ho Chi Minh city}
 \country{Vietnam} 
} 
\email{nvu@fit.hcmus.edu.vn}
\orcid{0000-0002-0594-4372}

\author{Tien N. Nguyen}
\affiliation{%
 \institution{Computer Science Department}
 \institution{University of Texas at Dallas}
 \city{Dallas}
 \state{Texas}
 \country{USA}} 
\email{tien.n.nguyen@utdallas.edu}
\orcid{0009-0006-7962-6090}

\begin{abstract}
In API testing, deriving logical constraints on API response bodies to be used as oracles is crucial in generating test cases and performing automated testing of RESTful APIs. However, existing approaches are restricted to dynamic analysis in which oracles are extracted via the execution of APIs as part of the system under test. 
In this paper, we propose a complementary LLM-based, static approach in which the constraints for API response bodies are mined from  API specifications. We leverage large language models (LLMs) to comprehend the API specifications, mine constraints for response bodies, and generate test cases. To reduce LLMs' hallucination, we apply an Observation-Confirmation (OC) scheme which uses initial prompts to contextualize constraints, allowing subsequent prompts to more accurately confirm their presence.
Our empirical results show that {\tool} with OC prompting achieves high precision in constraint mining with the average from 85.1\%--93.6\%. 
It also performs well in generating test cases from mined constraints, with a precision from 86.4\%--91.7\%. We also use the test cases generated by {\tool} to detect 46 mismatches between the API specification and actual response data for 19 real-world~APIs. Four of the mismatches were, in fact, reported in developers' forums.
\end{abstract}










\maketitle

\input{sections/intro}
\input{sections/motiv}
\input{sections/constraints-mining}

\input{sections/constraints-test-generation}

\input{sections/empirical}
\input{sections/results}

\input{sections/threats-to-validity}

\input{sections/discussion}

\input{sections/conclusion}

\section{Data Availability}

Our benchmark and code are publicly available at our website~\cite{website}.

\section*{Acknowledgments}
This work is funded by the University of Science, VNU-HCM, under Grant No. CNTT 2025-21.
Tien N. Nguyen was supported in part by the US National Science Foundation (NSF) grant CNS-2120386.


\newpage

\balance

\bibliographystyle{ACM-Reference-Format}

\bibliography{references-1, references-2, references}

\end{document}

%% file: sections/intro.tex
\section{Introduction}



By adhering to the principles of Representational State Transfer (REST), the RESTful APIs provide a standardized way for interoperability among components and software systems. RESTful API testing helps identify and resolve several issues, ensuring that APIs perform as expected~\cite{viglianisi2020resttestgen-2, MartinLopez2021Restest-2, arcuri2019restful, liu2022morest-2, alonso2022arte, atlidakis2019restler-2}. Among techniques for API testing, black-box testing uses the OpenAPI Specification (OAS) as a basis to generate test cases and data \cite{kat-2, karlsson2020quickrest-2, MartinLopez2021Restest-2}. The state-of-the-art API testing approaches are focused on status code~\cite{viglianisi2020resttestgen-2,kat-2,katalon, postman}, schema validation~\cite{viglianisi2020resttestgen-2,katalon,postman}, and rule extraction from OAS~\cite{kim2023enhancing-2}. Existing static code validation approaches define an oracle for a test case by validating whether the response status code matches the expected value. In contrast, {\em schema validation} ensures the correctness of the response data by checking it against a specified schema.

While status code and schema validation effectively cover aspects of data representation and status checking, they may overlook the {\em logical correctness and validity of the response data from the APIs}. For example, an API request for a customer older than 18 receiving a response for one younger than 18 would not be detected by just validating the status code or schema. 
%
The state-of-the-art approaches for mining constraints on API response bodies focus only on  {\em dynamic analysis} as the constraints are extracted from the execution data of the system under test (SUT). AGORA/AGORA+\cite{alonso2023agora, Alonso2025AGORA_plus} extends Daikon~\cite{ernst2007daikon}, a dynamic instrumenter, to detect invariants during the execution of the APIs within the SUTs.
This paper presents {\tool}, a novel LLM-based approach to mine the {\em constraints of API response bodies} from the API specification. To mine constraints, we leverage the ability of large language models (LLMs) to comprehend textual descriptions in API specifications. Constraints on API response bodies are inferred from different sources, including response properties, response schema, operations, and request parameters. We also harness LLMs’ proficiency to create test cases from the mined constraints to verify if the SUT correctly returns the content satisfying the mined constraints.~Moreover, we apply an {\em Observation-Confirmation} scheme by dividing the task of mining constraints into two phases: observation and confirmation. The initial prompt contextualizes the description of constraints, enabling the next prompt to more accurately decide their presence. Moreover, LLMs might have hallucinations.
Thus, we enhance {\tool} with two extra mechanisms. First, before requesting the LLM to make observations concerning constraints on parameters, we perform {\em a filtering process} to keep only the valid ones. Second, after generating test cases for mined constraints, we add a {\em semantic verifier} to verify those test cases against the examples specified in the OAS file. The idea is that such examples tend to be correct because they illustrate the descriptions on the data. For example, the OAS could give "March" as a valid month.
If such a correct example does not pass a test case generated by {\tool} based on the mined constraint(s), the test case must be incorrect, which is caused by incorrect constraint(s). Thus, our verifier will discard them, leading to improved precision.




We evaluate {\tool} on two datasets, AGORA/AGORA+ \cite{alonso2023agora}  and our own dataset collected from 8 real-world API services consisting of 65 endpoints and 90 operations \cite{website}. 
Our empirical results show that {\tool} achieves high precision in constraint mining with 93.6\% on average for the \tool dataset and 85.1\% for the AGORA/AGORA+ dataset. Precision in test generation for constraint validation is from 86.4\%--91.7\%. We also use its generated tests to automatically verify mined constraints and detect mismatches between the API specifications and actual execution of the SUTs. A detected mismatch indicates a fault in a SUT or that the SUT does not match its specification. We report 46 actual faults~found in 19 real-world applications, including 4 issues reported by users on GitLab Forum \cite{gitlab, gitlab_issue_time1, gitlab_issue_time2, gitlab_issue_time3}.
This paper makes the following contributions:

{\bf 1. {\tool}: An LLM-based method} to extract constraints~from API specifications and to generate tests to validate API responses.

{\bf 2. A manually-verified benchmark} with ground-truth constraints for API response bodies of 8 real-world API services is available~\cite{website}
for future research on API testing approaches.



{\bf 3. An extensive evaluation} showing {\tool}'s capabilities in constraint or oracle mining and test generation for API testing.

%% file: sections/motiv.tex
\vspace{-3pt}
\section{Motivation}

Let us consider an example with Stripe, an online payment service.
Figure~\ref{listing:stripe_spec_sample}(b) partially shows a description from the API specification (OAS), detailing the \texttt{GET} operation for retrieving charge records
within a time interval (lines 9-10), which is defined by the \texttt{gt} and \texttt{lt} parameters (i.e., the lower and upper bounds).
Users can also specify the customer for whom they wish to retrieve charging history (lines 15-17).
A successful request returns a response with a status code of \texttt{200} and a response body with data. The structure of the response data (in Figure~\ref{listing:stripe_spec_sample}(a))
consists~of a list of \code{charge} objects, each associated with various properties, e.g., amount, created timestamp, currency, and others.
For instance, a \texttt{GET} request to the \code{/v1/charges} endpoint with parameters like \code{`created[gt]=1679090500\&customer=cus\_idA'} returns a list of charges that satisfy the given conditions. An example response object is in Figure~\ref{listing:stripe_response_body_sample}. The response's content is~constrained by the actual input parameters in the request. Thus, a~complete testing process needs to verify the response content and the returned status code.
For example, a test case can check if the \code{created} field of a returned charge object falls within the interval and ensuring that the \code{customer} field matches the requested~ID.

\begin{Observation} [Constraints from input parameters]
\label{o1}
The response data is constrained by the input parameters from the request. When testing, in addition to verifying the status code, testers need to verify the response bodies returned from the APIs.
\end{Observation}

\input{sections/motiv-example}

\input{sections/fig-response}

In Figure~\ref{listing:stripe_spec_sample}(a), the descriptions on the properties of the returned
\code{charge} objects define certain constraints on the attributes. For example, the \code{amount} property must be positive, with a max value of eight digits. The \code{currency} attribute has a three-letter lowercase code.

\begin{Observation} [Constraints within response body]
\label{o2}
Natural language descriptions on operations express logical constraints on operations, their responses, formatting requirements, or value range limitations that must be validated during API testing.
\end{Observation}

The OAS file often includes examples that illustrate specific constraints on data or data types. For instance, in Figure~\ref{listing:stripe_spec_sample}(a), the file provides an example of \$999,999.99 as a positive number for the property \code{amount}. A mined constraint that does not match its corresponding example may be incorrect. To validate that, we utilize the generated test case for the constraint.


\begin{Observation} [Verification with examples]
\label{o3}
The illustrating examples in the description can be used to verify against the generated test cases, i.e., the validity of the mined constraints.
\end{Observation}


\input{sections/related} 

\section{Key Ideas}
\label{sec:ideas}



We design {\tool}, an LLM-based, static approach for mining the constraints of API response bodies and then generating test cases.


\begin{figure*}[ht]
  \centering
  \includegraphics[width=0.9\textwidth]{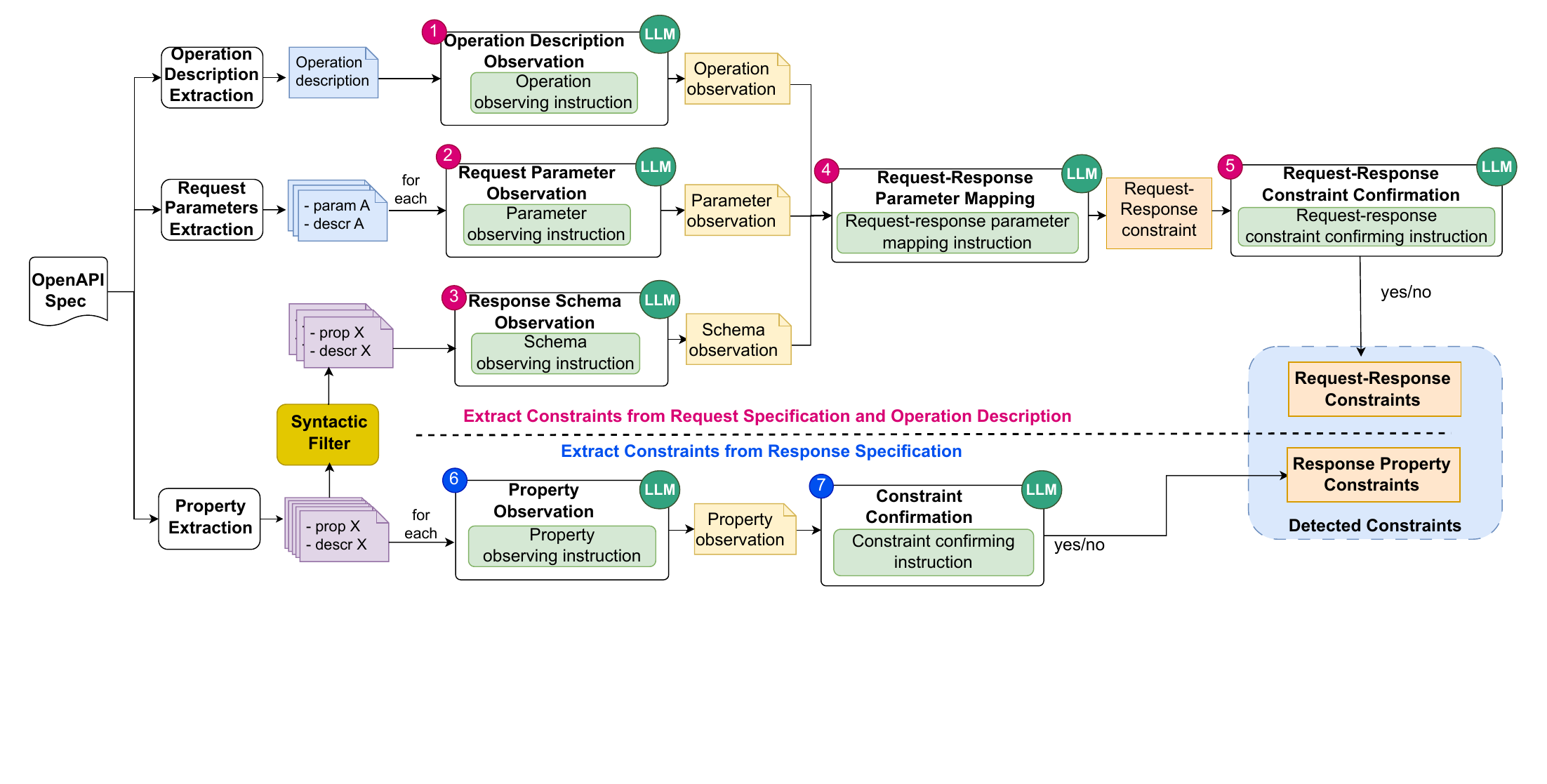}
  \vspace{-66pt}
  \caption{Mining Constraints in APIs' Response Bodies with LLMs}
  \label{fig:method-overview}
\end{figure*}

\vspace{4pt}
{\bf \em Key Idea 1 [Leveraging LLMs to Mine Constraints in APIs' response bodies].} {\tool} is a novel LLM-based static approach to mine the constraints for API's response bodies from the OAS. The constraints on the response data can be found in the specification in either the descriptions of the {\em operations} (e.g., line 5 of Figure~\ref{listing:stripe_spec_sample}(a)) or the descriptions of the {\em schema for the response data} (lines 2--24 in Figure~\ref{listing:stripe_spec_sample}(b)). For example, the description in Figure~\ref{listing:stripe_spec_sample} states that {\em `the charges are returned in a sorted order with the most recent ones appearing first'}. The schema for the returned values in Figure~\ref{listing:stripe_spec_sample} also provides us several constraints on the {\em parameters} (lines 5--6) as well as the description of the returned object (line 2). For example, the amount of a charge is a positive number with up to 8 digits and such value must be of integer. In brief, {\tool} is useful in scenarios where APIs and their execution are unavailable or unreliable. For example, this occurs when an API specification is available, but the APIs and their underlying systems are still under development.





\textbf{\bf \em Key Idea 2 [Observation-Confirmation scheme on LLMs for constraints discovery and test generation]}. For the task of extracting constraints from API specifications, we utilize the ability of LLMs to comprehend natural language descriptions found in API specifications for constraint mining on the APIs and their parameters. Our experiment showed that direct use of LLMs for constraint mining yields sub-optimal performance. To improve it, we apply an Observation-Confirmation scheme in which the initial result returned from the LLMs will be fed back to themselves in a confirmation prompt to provide better contexts on the constraints.






\vspace{1pt}
\textbf{\em Key Idea 3 [Generating test cases for the constraints on response bodies].} Our goal is to advance beyond current API testing techniques: we also generate test cases to evaluate the mined constraints. For instance, a test case is generated to verify that the list of charges returned by the API endpoint `v1/charges' is sorted in reverse chronological order. Another example includes generating a test case to validate the format/value of the returned charge~amount.



\vspace{1pt}
{\bf \em Key Idea 4 [Filter and Semantic Verifier].} Before asking the LLM to analyze constraints on parameters, we perform filtering to remove invalid constraints. After generating test cases based on the mined constraints, we introduce a semantic verifier to check these test cases against the examples in the OAS. The rationale is that these examples are typically accurate in the specification. For instance, an example of \$999,999.99 is used to illustrate a positive number for \code{amount}. If a test case fails to validate a given correct example, it suggests that the mined constraint may be~incorrect.




%% file: sections/motiv-example.tex
\begin{figure}[t]
\centering
\lstset{
		numbers=left,
		numberstyle= \tiny,
		keywordstyle= \color{blue!70},
		commentstyle= \color{red!50!green!50!blue!50},
		frame=shadowbox,
		rulesepcolor= \color{red!20!green!20!blue!20} ,
		xleftmargin=1.5em,xrightmargin=0em, aboveskip=1em,
		framexleftmargin=1.5em,
                numbersep= 5pt,
		language=Java,
    basicstyle=\scriptsize\ttfamily,
    numberstyle=\scriptsize\ttfamily,
    emphstyle=\bfseries,
                moredelim=**[is][\color{red}]{@}{@},
		escapeinside= {(*@}{@*)}
	}
\begin{lstlisting}[style=yaml]    
(a) charge:
    description: The 'charge' object represents an attempt to move money into your account.
    properties:
        amount:
          description: A positive integer can be up to eight digits.
          type: integer
          example: 99999999
        created:
          description: Time at which the object was created. Measured in seconds since the Unix epoch.
          type: integer
        currency:
          description: Three lowercase letters.
          type: string
          example: usd
        customer:
          description: ID of the customer this charge is for if existed.
          type: string ...
\end{lstlisting}
\vspace{-9pt}
\begin{lstlisting}[style=yaml]
(b) paths:
  /v1/charges:
    get:
      description: Returns a list of charges you have created. The charges are returned in sorted order ...
      parameters:
          name: created
          description: Only return charges that were created during the given date interval.
          schema:
            anyOf:
              - properties:
                  gt (integer)
                  lt (integer)
          name: customer
          description: Only return charges for the customer specified by this customer ID.
          schema:
            type: string ...
\end{lstlisting}
\vspace{-18pt}
\caption{(a) Schema of a response for `charge' API in the project Stripe described in a OAS file and (b) a simplified Description for the GET charges API operation from Stripe.}
\label{listing:stripe_spec_sample}
\end{figure}

%% file: sections/fig-response.tex
\begin{figure}[t]
	\centering
	\lstset{
		numbers=left,
		numberstyle= \tiny,
		keywordstyle= \color{blue!70},
		commentstyle= \color{red!50!green!50!blue!50},
		frame=shadowbox,
		rulesepcolor= \color{red!20!green!20!blue!20} ,
		xleftmargin=1.5em,xrightmargin=0em, aboveskip=1em,
		framexleftmargin=1.5em,
                numbersep= 5pt,
		language=java,
    basicstyle=\scriptsize\ttfamily,
    numberstyle=\scriptsize\ttfamily,
    emphstyle=\bfseries,
	}
\begin{lstlisting}[]
{
  "id": "ch...15",
  "object": "charge",
  "customer": "cus_id",
  "amount": 1099,
  "created": 1679090539,
  "currency": "usd",...
}
\end{lstlisting}
\vspace{-18pt}
\caption{A Response's Body from a \code{GET} Request for Stripe.}
\label{listing:stripe_response_body_sample}
\end{figure}

%% file: sections/related.tex
\section{Related Work}


Recent surveys on API testing~\cite{kim2022automated-2,golmohammadi2022testing-2, ehsan2022restful, martin2022online-2,  sharma2018automated,marculescu2022faults-2,martin2021black-2} reveal a trend towards automation adoption. AI/ML are used to enhance various aspects of API testing including of generating  test cases~\cite{viglianisi2020resttestgen-2, MartinLopez2021Restest-2, atlidakis2019restler-2}, realistic test inputs~\cite{alonso2022arte}, and identify defects early in the development~\cite{liu2022morest-2, sahin2021discrete-2, arcuri2020automated, arcuri2021enhancing,arcuri2020handling, zhang2021adaptive-2, zhang2021enhancing-2,zhang2019resource-2,zhang2021resource-2}. In API test case generation, validation typically falls into two main categories.

\textbf{Status Code Validation:}
Each HTTP request is returned with a response containing a status~code and data. The status code, a 3-digit integer, indicates the outcome of the HTTP request. \texttt{2xx} codes signify a successful request. Conversely, \texttt{4xx} codes indicate errors, e.g., a bad syntax request or invalid input values. For instance, in testing the API \code{'GET/user\_information'} with various valid and invalid user IDs, testers would expect the API to return status codes \texttt{200} or \texttt{404}. This validation method is used in automated tools, e.g., Postman~\cite{postman}, Katalon~\cite{katalon}, RestTestGen~\cite{viglianisi2020resttestgen-2}, and KAT~\cite{kat-2}.



\vspace{2pt}
\textbf{Schema Validation:}
This verifies the presence of all required properties and the consistency of property data types with specifications. 
Tools, e.g., RestTestGen~\cite{viglianisi2020resttestgen-2}, leverage external libraries, such as ‘swagger-schema-validator’, to facilitate schema validation. 

While status code and schema validation cover data representation and status checking, they {\em overlook the logical correctness and validity in the API responses}. For instance, if an API request for a charge in 2025 returns one from 2024, or if the amount is negative, these would not be detected by validating the status code or schema.


AGORA/AGORA+~\cite{alonso2023agora,Alonso2025AGORA_plus} extends Daikon~\cite{ernst2007daikon} to infer the invariants from the execution data of the SUT using APIs. SATORI~\cite{alonso2025satori} uses LLMs to infer API's expected behavior by analyzing the properties of the response fields of its operations. It does not 
have Observation-Confirmation (OC) scheme as in {\tool}.

KAT~\cite{kat-2} fully automates API testing using GPT and an input OpenAPI specification, building dependency graphs and generating test scripts. 
Kim {\em et al.}~\cite{kim2024leveraging-2} applied GPT to enrich specifications with rule explanations and example inputs.
Before LLMs, ARTE \cite{alonso2022arte} used NLP to generate API test inputs. Morest~\cite{liu2022morest-2} proposed model-based RESTful API testing with a dynamic Property Graph, improving code coverage and bug detection. RESTler \cite{atlidakis2019restler-2} introduced stateful fuzzing of REST APIs, inferring dependencies and generating tests guided by service responses. LlamaRestTest~\cite{kim2025LlamaRestTest} employs fine-tuned and quantized Llama3-8B model using mined datasets of REST API example values and inter-parameter dependencies, to generate realistic test inputs and uncover inter-parameter dependencies.




%% file: sections/constraints-mining.tex
\section{Mining Constraints for API Response Bodies}
\label{sec:constraint_mining}

The specification for an API endpoint includes two main parts: the {\em request specification} (the {\bf input} of the API) and {\em response schema specification} (the {\bf output} of the API). (1) A request specification determines how to call an API endpoint, detailing the required {\em input parameters}, their roles, and the parameters within the request body along with their respective roles. (2) The response schema specification provides instructions on the {\em response body}, including each property in the response data, its description, datatype, nullability, and other details. These parts are the target for constraint mining. 



\input{sections/mining-request-parameters}

\vspace{-2pt}
\subsection{Constraints from Response Specifications}
\label{subsec:response-param}

\begin{algorithm}[t]
\small
\caption{Extract Constraints from Response Specifications}
\label{alg:constraint_extraction}
\begin{flushleft}
\textbf{Input:} API\_spec(dict): obj of entire API specification.\\
\qquad response\_schema(dict): schema obj of a certain response.\\
\qquad knowledge\_base(dict): gained knowledge.\\
\textbf{Output:} list of constraint properties. \\
\end{flushleft}
\begin{algorithmic}[1]
\STATE \textbf{function} getConstraintInsideResponseSchema(API\_spec, response\_ schema, knowledge\_base)
\STATE \quad constraint\_properties $\leftarrow$ []
\STATE \quad \textbf{for each} prop \textbf{in} resp\_schema \textbf{do}
\STATE \quad\quad \texttt{desc} $\leftarrow$ resp\_schema[prop]["description"]
\STATE \hspace*{1.2em} +resp\_schema[prop]["type"]+resp\_schema[prop]["format"]
\STATE \hspace*{1.2em} +resp\_schema[prop]["example"]
\STATE \quad \quad \textbf{if} desc = NULL \textbf{then}
\STATE \quad \quad \quad desc $\leftarrow$ exactMatchProp(API\_spec, prop)
\STATE \quad \quad \quad \textbf{if} desc = NULL \textbf{then}
\STATE \quad \quad \quad \quad \textbf{continue} \# Skip this property

\STATE \quad \quad \textbf{if} prop \textbf{in} knowledge\_base \textbf{then}
\STATE \quad \quad \quad \textbf{if} knowledge\_base[prop] = TRUE \textbf{then}
\STATE \quad \quad \quad \quad constraint\_properties.append(prop)
\STATE \quad \quad \textbf{else}
\STATE \quad \quad \quad datatype $\leftarrow$ response\_schema[prop]["datatype"]

\STATE \quad \quad \quad prop\_obser $\leftarrow$ LLM().propertyObser(prop, datatype, desc)

\STATE \quad \quad \quad constraint\_confirmation $\leftarrow$ LLM().constraint\_confirmation (prop, datatype, desc, prop\_obser)

\STATE \quad \quad \quad \textbf{if} constraint\_confirmation = TRUE \textbf{then}
\STATE \quad \quad \quad \quad constraint\_properties.append(prop)

\STATE \quad \quad \quad \# Add this property to knowledge base
\STATE \quad \quad \quad knowledge\_base[prop] $\leftarrow$ constraint\_confirmation
\STATE \quad \textbf{return} constraint\_properties
\STATE \textbf{end function}
\end{algorithmic}
\end{algorithm}

We examine descriptions, data formats, and examples in the response schema specification, as they may provide constraints for a property. Each endpoint includes a response schema that guides clients on parsing returned data, structuring the response object with properties, descriptions, data formats, and examples (Figure~\ref{listing:stripe_response_body_sample}).  

Algorithm~\ref{alg:constraint_extraction} shows our constraint-mining algorithm for response specifications. It first extracts descriptions, data formats, and examples for a property (lines 4-10). If such information exists, we check the knowledge base of LLM-identified constraints to avoid redundant queries (lines 11-13). If found, we reuse the stored data; otherwise, we prompt the LLM to extract constraints (lines 15-21). 


\vspace{2pt}
{\bf Observation-Confirmation Strategy}. This involves two phases: {\em observation} (line 16) and {\em confirmation} (line 17). Our experiments reveal that when descriptions lack detail for constraint extraction, the LLM may resort to fabricating details, a phenomenon known as hallucination. Drawing inspiration from the Chain-of-thought~\cite{chain_of_thought}, we divide the task of extracting constraints into two phases of {\bf observation} and {\bf confirmation}, the initial prompt better contextualizes the description of constraints, enabling the subsequent prompt to more accurately determine a constraint. 

In the observation phase (Block \circlednum[response]{6}), the LLM is prompted to identify constraints from the description. For example, given a~property \code{date} (type: \code{string}) described as {\em “ISO date: the literal date of the holiday”}, the LLM might infer: {\em "The date must follow the YYYY-MM-DD format for validity, ensuring consistency within the API.”} 

Next, the observation is fed into the Constraint Confirmation Prompt (Block \circlednum[response]{7}), where the LLM validates whether the extracted constraint provides enough detail for script generation. This step, similar to Figure~\ref{prompt:MAPPING_CONFIRMATION}, ensures that constraints specify values, ranges, or formats. If confirmed, the constraint is marked as a Response Property constraint and stored in the knowledge base.

%% file: sections/mining-request-parameters.tex
\subsection{Constraints from Request Specifications}
\label{subsec:request-spec}

Figure~\ref{fig:method-overview} describes the process of mining constraints. Constraints for response bodies may exist in descriptions of operations, parameters, and other information defined in the request specifications.

\input{prompts/mapping_prompt}

\input{prompts/constraint_confirmation_prompt}
\subsubsection{Request-Response Constraint Mapping}
The response data may contain a property that reflects a constraint available in request parameters. Thus, we {\em identify pairs consisting of a request parameter that includes a constraint and a response data property that reflects this constraint} (Figure~\ref{prompt:MAPPING_PROMPT}). For instance, to verify a "customerID" constraint, the response data should contain a property that identifies the customer. Thus, the required pair is <"customerID", "customer">. If the response data does not have a field to represent a constraint, that constraint is disregarded as it cannot be validated.

\subsubsection{Detailed Process}
The process of extracting constraints from request parameters encompasses four steps: (1) description extraction, (2) {\bf observation} (\circlednum[request]{1}-\circlednum[request]{3} from Figure~\ref{fig:method-overview}), (3) Request-Response constraint mapping (\circlednum[request]{4}), and (4) constraint {\bf confirmation} (\circlednum[request]{5}).

\begin{algorithm}[t]
\small
\caption{Extract Constraints from Request Specifications}
\begin{flushleft}
\textbf{Input:} API\_spec(dict): obj of entire API specification. \\
\qquad req\_spec(dict): specification obj of a certain request. \\
\qquad resp\_schema(dict): schema obj of the associated response. \\
\textbf{Output:} list of request-response constraint properties
\end{flushleft}

\label{alg:request_response_constraint_extraction}
\begin{algorithmic}[1]
\STATE \textbf{function} getConstrFromReqParas(API\_spec, req\_spec, resp\_schema)
\STATE \quad reqRespConstraints $\leftarrow$ []

\STATE \quad respSchemaObser $\leftarrow$ LLM().respSchemaObser(resp\_schema)

\STATE \quad operationObservation $\leftarrow$ LLM().operationObser(desc)
\STATE \quad \textbf{for each} param \textbf{in} req\_spec \textbf{do}
\STATE \quad \quad desc $\leftarrow$ req\_spec[param]["desc"]
\STATE \quad \quad \textbf{if} req\_spec[param]["desc"] = NULL \textbf{then}
\STATE \quad \quad \quad desc $\leftarrow$ findExactMatchParameter(API\_spec, param)\\
\STATE \quad \quad \quad \textbf{if} desc = NULL \textbf{then}
\STATE \quad \quad \quad \quad \textbf{continue} \# Skip this param

\STATE \quad \quad \# Use LLM to find constraint for this param
\STATE \quad \quad paramObser $\leftarrow$ LLM().parameterObser(param, desc)

\STATE \quad \quad answer, corrProp, explain $\leftarrow$ LLM().reqRespMapping\\ \quad \quad \quad (param, desc, paramObser, respSchemaObser)

\STATE \quad \quad \textbf{if} answer = TRUE \textbf{then}
\STATE \quad \quad \quad confirmation $\leftarrow$ LLM().confirmReqRespMapping\\(param, corrProp, explain)
\STATE \quad \quad \quad \textbf{if} confirmation = TRUE \textbf{then}
\STATE \quad \quad \quad \quad reqRespConstraints.append((param, corrProp))

\STATE \quad \textbf{return} reqRespConstraints
\STATE \textbf{end function}
\end{algorithmic}

\end{algorithm}

Algorithm~\ref{alg:request_response_constraint_extraction} provides the details.
For each API request, a prompt is made to gather observations on the response schema and operations associated with this request (lines 3--4, Algorithm~\ref{alg:request_response_constraint_extraction}). With each request parameter (line 5), we engage the LLM to offer observations on the parameter using its description (line 12, Algorithm~\ref{alg:request_response_constraint_extraction}).
These parameter observations support the next step: Request-Response parameter mapping (Block \circlednum[request]{4}, line~13). We guide the LLM through a two-step reasoning process using a tailored prompt (Figure~\ref{prompt:MAPPING_PROMPT}). The LLM receives a brief description of the request parameter, including its details and observations (\circlednum[request]{2}), ensuring an understanding of its intent and constraints. We then present the response schema, observations from Block \circlednum[request]{3}, and ask the LLM to identify a matching property. A match occurs when the request parameter filters response data or both share the same~meaning. 

From this two-step reasoning, we require the LLM to answer three questions: (1) Is there a matching property in the response schema? (2) What is this property? (3) How does the request parameter influence this property?

%

If the answer to (1) is false, i.e., no property reflects this request parameter, we disregard this parameter. Otherwise, we proceed to the final prompt (Figure~\ref{prompt:MAPPING_CONFIRMATION}), which is the confirmation of the mapping (lines 14--17, Block \circlednum[request]{5}). We present the pair of the request parameter and the matched property from Block \circlednum[request]{4}, and ask the LLM to confirm the accuracy of this mapping (line~15). This step aims to minimize the occurrence of false positives. Finally, the validated pairs of request parameters and response properties are stored as Request-Response constraints (lines 16--17). 

%% file: prompts/mapping_prompt.tex
\begin{figure}[t]
\begin{lstlisting}[style=pythonStyle]
PARAMETER_SCHEMA_MAPPING_PROMPT = '''Given a request parameter and an API response schema, check if there is a matching property in the API response schema.
Request parameter for {method} - {endpoint}:
{parameter}: {description}

Follow these steps to find the matching property: {Instructions for chain-of-thought steps}

Schema specification {schema}: {schema_observation}
Confirm if the request parameter has a matching property in the response schema: ...
Identify the corresponding property name of the provided request parameter in the schema. ...
If a matching property exists, explain it in this format:...
\end{lstlisting}
\vspace{-18pt}
\caption{Request-Response Parameter Mapping Observation}
\label{prompt:MAPPING_PROMPT}
\end{figure}

\begin{figure}[ht]
\centering
\begin{lstlisting}[style=pythonStyle]
MAPPING_CONFIRMATION = '''{System prompt}
The request parameter's information:
- Operation: {method}
- Parameter: {parameter_name}
- Description: {description}
The corresponding property's information:
- Resource: {schema}
- Corresponding property: {corresponding_property}
{Instructions for chain-of-thought steps}
Answer format: ...'''
\end{lstlisting}
\vspace{-18pt}
\caption{Request-Response Constraint Confirmation}
\label{prompt:MAPPING_CONFIRMATION}




\end{figure}

%% file: sections/constraints-test-generation.tex
\section{Constraint Test Generation}
\label{sec:test-generation}

\begin{figure}[t]
  \centering
  \includegraphics[width=3.2in]{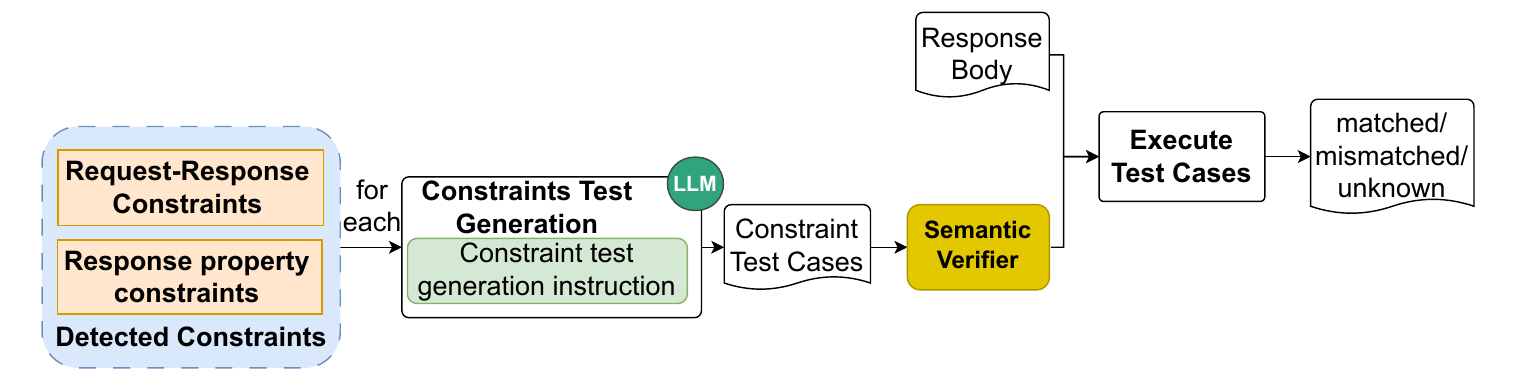}
 \vspace{-9pt}
  \caption{Constraint Test Generation in {\tool}}
  \label{fig:test-gen}
\end{figure}

\input{prompts/test-gen-prompt}
Taking the mined constraints, {\tool} generates test cases for automated verification of API response bodies against their constraints. This process (Figure~\ref{fig:test-gen}) uses LLMs to convert Request-Response and Response Property constraints into test cases. Each resulting test case is a validation function that takes response data and request parameters as input, and returns whether the response data satisfies their corresponding mined constraints.
We also incorporate a semantic verifier that cross-checks test cases against examples in the OAS file.  If a test case fails to validate a given correct example, it suggests that the mined constraint may be incorrect.




\vspace{3pt}
{\bf Request-Response Constraints Testing.} Such cases verify dependencies between a request parameter and its response property. The prompt 
to generate such test cases (Figure~\ref{prompt:CONSTRAINT_TEST_GEN_PROMPT}) requires four inputs:  API input parameter, mined constraints,  its property, and response schema.  
%
To ensure robustness, {\tool} guides the LLM with predefined rules: using a \code{try-catch} block for error handling, excluding examples, and following a standardized function template. These rules ensure consistency and focus on constraint~verification.





\vspace{3pt}
{\bf Response Property Constraints Testing.} Test cases for Response Property constraints are used to verify response data against their constraints. To create a test case~for this type of constraint, {\tool} uses the prompt depicted in Figure~\ref{prompt:CONSTRAINT_TEST_GEN_PROMPT}. This prompt requires three inputs: the response property, its mined constraints, and response schema. The response property and its mined constraints provide the necessary information for generating validation code, while the response data schema defines the structure and type of the expected data, allowing the LLM to parse response data correctly. 


%% file: prompts/test-gen-prompt.tex

\begin{figure}[t]
\centering
\begin{lstlisting}[style=pythonStyle]
CONSTRAINT_TEST_GEN_PROMPT = '''
Generate a Python script to check if a property in a REST API's response meets specified constraints and rules.
Constraint description:
- Constraint from request parameter: {parameter}
- Constraint description: {constraint_description}
API response schema: {response_schema_specification}
The property of the request parameter in API response:
- "{property}": "{prop_description}"
Generate a Python script to verify '{property}' in response. 
Rules: {Rules for test gen}
Format the script as shown: ...'''
\end{lstlisting}
\vspace{-18pt}
\caption{Constraint Test Generation Prompts}
\label{prompt:CONSTRAINT_TEST_GEN_PROMPT}
\end{figure}

%% file: sections/empirical.tex
\section{Empirical Evaluation}
\label{sec:eval}


To evaluate {\tool}, we seek to answer the following questions:


\vspace{2pt}

\textbf{RQ1. [Constraints Mining Accuracy]}. How well does {\tool} mine the constraints to be used as oracles for verifying API responses compared with the dynamic approach AGORA+~\cite{Alonso2025AGORA_plus}?

\vspace{1pt}
\textbf{RQ2. [Constraints Test Generation Accuracy]}. How well does {\tool} generate test cases from the mined constraints?

\vspace{1pt}
\textbf{RQ3. [Accuracy in Test Generation from Correctly Mined Constraints]}. How accurate are test cases from correct constraints?


\textbf{RQ4. [Usefulness in Response Body Testing]}. How useful is {\tool} in detecting mismatches between the specification and its working APIs?

\textbf{RQ5. [Ablation Study]}. How much do Observation-Confirmation and Semantic Verifier contribute to \tool's performance?

\input{sections/dataset-2}

%% file: sections/dataset-2.tex


\vspace{2pt}

\noindent {\bf Datasets.} We used the following two datasets:

{\em 1. AGORA+ dataset~\cite{Alonso2025AGORA_plus}} has 7 API services with 11 operations.

\vspace{1pt}
{\em 2. RBCTest dataset~\cite{rbctest-dataset}.}
We curate the dataset from eight real-world services from GitLab and Stripe, comprising 65 endpoints and 90 operations from `Canada Holidays'~\cite{kat-2}, GitLab-services~\cite{yamamoto2021efficient-2, wu2022combinatorial-2, atlidakis2019restler-2, karlsson2020quickrest-2, lin2022forest, kat-2}, and `Stripe'~\cite{kat-2,martin2022online-2,mirabella2021deep_stripe-2}. These services were selected for several reasons: (1) They feature complex request-response structures across diverse business domains. (2) They have been used in API testing. (3) Their active status allows API calls to collect real response data. (4) Their specifications vary in documentation quality, enabling evaluation across different levels of completeness. Stripe’s test mode offers limited endpoints and often returns deeply nested, incomplete schemas that miss inputs, hindering validation, so we restricted our study to 8 endpoints without nested schemas.


For both datasets, we manually examine the specifications and build the ground truth of correct constraints, i.e., test oracles. During examination, we systematically used the following criteria to look for and derive the correct and sufficient constraints:

{\em (1) Request-Response Property constraint}: This type of constraint on a request parameter must correspond to a property in the response data that reflects this constraint. 

{\em (2) Range-of-Value constraint}: The type of constraint must specify all possible values or provide a specific data range.

{\em (3) Data Format constraint}: This must describe the constraints on data format or refer to widely used formats (e.g., ISO, Unix).

{\em (4) Computed Value constraint}:
  The constraint describes how the value of a field is derived from other fields via computation.

{\em (5) Name-based constraint}: This primarily focuses on the constraints held within field names like \code{id, code, email, url, uri}, etc.

{\color{custom-blue}{
We further referred to the oracles available in the AGORA+ dataset to define and report the oracle categories in the {\tool} dataset (Table~\ref{tab:gt-distribute}). The first category is the input/output constraints, which describe the logical relations between the input parameters of APIs and the output properties in their response bodies. The remaining categories contain only the constraints/oracles on the output properties in the response bodies. Specifically, the second category contains the constraints on the individual variables in the output.
This category contains {\em unary} constraints belonging to several different data types and templates ({\em string, boolean, float, time, url, email, etc}). The third category 
contains the {\em $N$-ary (binary or more)} atomic constraints, i.e., involving two or more variables (Table~\ref{tab:gt-distribute}). The atomic constraints include comparisons, membership/type checks, and predicate calls without any \code{AND}, \code{OR}, or universal/existential quantifiers (negations, if any, are pushed inside as part of the atomic). The last category is the super set of 
the three preceding categories. A {\em composite} constraint is expressed as a conjunctive/disjunctive form of the two or more preceding categories, i.e., logical \code{AND}/\code{OR} operations are applied among these~categories.
}}


\input{tables/gt-distribute}

%% file: tables/gt-distribute.tex
\begin{table}[]

\tabcolsep 2pt
\caption{Categories of Oracles in the \underline{{\tool} dataset}}
\vspace{-5pt}
\small
\label{tab:gt-distribute}
\footnotesize
\
\begin{tabular}{|l|p{0.35\textwidth}|}
\hline
\textbf{Category} & \textbf{Test Oracles and Examples} \\
\hline
\textbf{Input-Output} & Constraints involving the relation between input parameters and output properties. 

E.g.: \code{input.id\_after > return.id} \\
\hline
\textbf{Single-variable} & DefaultValue, isUrl, isDate, isDateTime, String\_Specific\_Length, Template-Literals, ArrayTypeOracle\_\{isString, isBoolean, isNumber\}, Value-In-Range, Value-In-Set, isBoolean, isNumber, isUnixTime, Array\_Specific\_Sizes, IsTime, isEmail, etc. \\
\hline
\textbf{N-ary atomic} & Constraints involving two or more variables.

E.g.: \code{sellingTotal = total + margins + markup + totalFees - discounts} \\
\hline 
\textbf{Composite} & Constraints with multiple I/O, single-variable, or atomic ones. 

E.g.: \code{input.created\_after} < \code{output.created\_at} and \code{output.created\_at} must isDateTime \\
\hline
\end{tabular}

\end{table}

%% file: sections/results.tex
\section{Experimental Results}

\input{sections/mining-eval}

\input{sections/complete-eval-new}

\input{sections/test-gen-only-eval}

\input{sections/response-body-testing}

\input{sections/ablation}

\input{sections/cost-analysis-2}

%% file: sections/mining-eval.tex
\subsection{Constraint Mining Accuracy (RQ1)}
\label{sec:rq1}

\subsubsection{Procedure} 
\label{sec:exp-methodology}

\input{sections/tab-constraint-invariant}

{\color{custom-blue}{
For the AGORA/AGORA+ dataset, we ran {\tool} with GPT-4o under the temperature of zero and top\_p=0.95 to obtain constraints and manually evaluated them against the ground truth. For the AGORA+ tool, we directly used the results reported in its TOSEM paper~\cite{Alonso2025AGORA_plus}. For the {\tool} dataset, we ran our tool in the same setting as with the other dataset. For the AGORA+ tool, we checked its publicly available repository and used RESTest~\cite{MartinLopez2021Restest-2} configuration set by AGORA+'s authors to generate 50 request-response pairs following their online instruction. Using this configuration (rather than the default RESTest configuration) ensures the realistic pairs for AGORA+ in {\tool}'s dataset. We chose 50 pairs since its authors reported the best tradeoff between performance and running time at this setting compared to other pair counts.


}}

%


Table~\ref{tab:example_both_methods} shows samples of constraints identified by {\tool} and invariants generated by AGORA+, illustrating their differences and limited compatibility. To enable meaningful comparison, we adopt a variable-based matching approach. In this context, a variable refers to any request parameter, response property, or intermediate value defined within the API specification. 
In AGORA+, each invariant is associated with a set of variables, as in Table~\ref{tab:example_both_methods}, and a given set of variables may be linked to one or more invariants (e.g. Invariants 3-4). In {\tool}, each constraint refers to one or more variables.     



If a constraint and an invariant involve disjoint sets of variables, we treat them as being uniquely identified by their respective approaches. If a {\tool}'s constraint and an AGORA+'s invariant refer to the same set of variables, we consider them to be \textit{overlapping}. To evaluate these overlapping pairs, we manually inspect both the constraint and the invariant, and then construct test cases to empirically validate or invalidate them. For example, Constraint 6 in Table~\ref{tab:example_both_methods}, we generate test cases with zero and ten guests per room. If both are found to be correct, we compare to see which~one is stricter (i.e., better). If one is correct and the other incorrect, we consider the correct one as better. We exclude from our comparison any pairs in which both the constraint and invariant are found to be incorrect. For metrics, we used true positives (TP), false positives (FP), false negatives (FN), and precision $P$ = $\frac{\text{TP}}{\text{TP} + \text{FP}}$, recall~$R$=$\frac{\text{TP}}{\text{TP} + \text{FN}}$. We repeated each experiment five times and reported {\em the mean~results}.








\subsubsection{Results on the AGORA+ dataset}
\label{sec:rq1-agoradataset}

\input{tables/rq1_constraints_mining_agora_dataset}

As seen in Table~\ref{tab:rq1-mining-agora-dataset}, \tool
identified an average of 576.8 constraints out of 616, with an average of 530.2 TPs, resulting in a precision of 85.1\% and a recall of 83.7\%. AGORA+ detected 953 invariants, and 473 invariants after variable-based grouping. Of these, 358 are TPs, resulting in 75.7\% precision.

\subsubsection{Results on the {\tool} dataset}

\input{tables/rq1_constraints_mining_ours_dataset}

We manually reviewed API specifications and noted 986 correct constraints (Table~\ref{tab:rq1-constraint-mining-ours-dataset}). Our tool identified {\em an average} of 746.8 constraints with a precision of 93.6\% and a recall of 70.6\%. All standard deviations are below~5\%.

\vspace{-2pt}
\subsubsection{Overlapping Analysis between {\tool} and AGORA+ results}
\label{sec:overlap}


For the AGORA+ dataset (Table~\ref{tab:rq1-mining-agora-dataset}), among overlapping ones, 328.4 constraints were uniquely detected by \tool, while 156.2 invariants were uniquely identified by AGORA+. Among the shared ones, {\em on average}, 45.4 from {\tool} are better while 23.2 from AGORA+ are better, indicating that {\tool} and AGORA+ are complementary and also share detected constraints. 

For the {\tool} dataset, regarding the overlapping results in Table~\ref{tab:rq1-constraint-mining-ours-dataset}, {\em {\tool} and AGORA+ are also complementary}, as {\tool} produces 143.4 better oracles with 290.0 uniquely detected ones, and AGORA+ has 28.8 better oracles with 1,035.2 unique ones.

An example of such complementary nature is the Spotify API. Its documentation states that three predefined image sizes (640, 300, 64) are served, but also notes that “the exact size may vary.” Here, AGORA+ detected three explicit values, while {\tool} identified the parameter as an integer, complementing each other’s findings.

\vspace{-1pt}
\subsubsection{Qualitative Analysis and Comparison} 

We examined the cases that {\tool} can detect, yet AGORA+ struggled. {\em First, the constraints uniquely detected by {\tool}} occurred in scenarios where the API specification explicitly includes descriptions of parameters, operations and response properties. AGORA+ relies on diverse runtime API responses, limiting its effectiveness when those responses lack diversity. For instance, in the \code{GET /api/v1/pro\-vinces} endpoint of the Canada Holidays API, AGORA+ inferred only that \code{LENGTH(return.id)==} 2, whereas {\tool} mined from the speci\-fication that \code{provinces.id} must be one of the enumerated two-letter Canadian province abbreviations (e.g., \code{"BC"}, \code{"ON"}, etc.). Similarly, for the attribute \code{currency} in \code{AmadeusHotel\_getMultiHotelOffers}, AGORA+ detected that the string length is three characters, however, {\tool} mined from the specification that the string must belong to the list of IATA codes (e.g., USD, JPY). Second, the specification describes a constraint in a more complete ranges of values or formulas than the specific values observed at the runtime. For example, in the Amadeus Hotel API, the \code{roomQuantity} value is specified as \textit{"an integer between 1 and 9."} {\tool} correctly identified this, whereas AGORA+ provided a general invariant of "Numeric". As another example, in the Amadeus Hotel API, the \code{sellingTotal} is defined as \textit{"= Total + margins + markup + totalFees - discounts."} While AGORA+ simply detected \code{sellingTotal} as "Numeric", {\tool}'s constraints mined from specifications are more specific and correct. {\tool} also infers better string templates, such as the \code{RunTime} attribute (e.g., \verb|^\d+\smin$|) in \code{OMDB\_byIdOrTitle}. Finally, {\tool} also mines better the constraints on the rarely exercised attributes, such as \code{template\_repository} in \code{Github\_createOrganiza\-tion\-Repository}.

We also examine the cases in which AGORA+ performed well while {\tool} struggled. These cases typically involve invariants that validate URL formats, handle fixed-length tokens, and capture inter-attribute relationships—areas where {\tool} often treats such attributes as plain strings without information. The analysis reveals that \tool is limited in these scenarios since it relies primarily on shallow attribute descriptions, typically consisting of the attribute name and a simple type declaration \code{(\{type:string\})}. For example, when an attribute such as \code{return.self} is described only with a type, \tool is unable to infer underlying constraints.

In contrast, AGORA+ leverages richer execution context and semantic information to detect the kinds of value-specific constraints. These constraints that AGORA+ detected better belong to three main types. {\em Type-based constraints} ensure that an attribute conforms to a specific data type, such as verifying that \code{return.owner.gists\_url} or \code{return.self} is of type \code{Url}. {\em Value-based constraints} impose conditions on numeric or string values, for instance, enforcing that \code{LENGTH(return.runners\_token) == 29} or that \code{return.stats.deletions >= 0} and does not exceed \code{return.stats.total}. Finally, {\em inter-attribute constraints} capture relationships among multiple fields, including equality \code{(return.owner.gists\_url} == \code{return.organization.gists\_url)} and substring relationships (e.g., \code{return.owner.url} being a substring of both \code{return.owner.followers\_url} and \code{return.owner.following\_url}).

\input{sections/detected-constraint-types-2}

%% file: sections/tab-constraint-invariant.tex
\begin{table}[t]
\centering
\footnotesize
\tabcolsep 2.2pt
\caption{{\tool}'s Constraints and AGORA+'s Invariants}
\label{tab:example_both_methods}
\vspace{-6pt}
\begin{tabular}{|c|c|c|}
\hline
\textbf{ID} & \textbf{Constraints}                                                                                                         & \textbf{Invariants}                                   \\ \hline
\textbf{1}  & \begin{tabular}[c]{@{}c@{}}the number of returned items\\ has to be less than or equal\\ to the requested limit\end{tabular} & input.limit \textgreater{}= size(return.items{[}{]})  \\ \hline
\textbf{2}  & --                                                                                                                            & return.total \textgreater{}= size(return.items{[}{]}) \\ \hline
\textbf{3}  & \multirow{2}{*}{\begin{tabular}[c]{@{}c@{}}return.total is an integer larger\\ than or equal  to 1\end{tabular}}             & return.total \textgreater{}= 1                        \\ \cline{1-1} \cline{3-3} 
\textbf{4}  &                                                                                                                              & return.total is Integer                               \\ \hline
\textbf{5}  & --                                                                                                                            & input.market is a substring of return.href            \\ \hline
\textbf{6}  & \begin{tabular}[c]{@{}c@{}}number of adult\\ guests (1-9) per room\end{tabular}                                              & return.adults is Integer                              \\ \hline
\textbf{7}  & -- & return.ci\_forward\_deployment\_enabled = true
\\ \hline
\textbf{8}  & An amount is a positive integer & -- \\ 
& can be up to eight digits                  & --                               \\ \hline
\textbf{9}  & -- & return.milestone = null
\\ \hline
\textbf{10}  & -- & return.labels[] == []
\\ \hline
\textbf{11}  & -- & return.statistics.wiki\_size one of \{ 0, 41943 \}
\\ \hline
\end{tabular}
\end{table}

%% file: tables/rq1_constraints_mining_agora_dataset.tex
\begin{table}[t]
\centering
\tabcolsep 0.8pt
\caption{Constraints Mining on the \underline{AGORA+ dataset} (RQ1)}
\vspace{-6pt}
\small
\label{tab:rq1-mining-agora-dataset}
\footnotesize
\begin{tabular}{|l|rrrrrr|rrrr|rrr|rr|}
\hline
\multicolumn{1}{|c|}{\multirow{2}{*}{\textbf{API}}} & \multicolumn{6}{c|}{\textbf{{\tool}}} & \multicolumn{4}{c|}{\textbf{AGORA+}} & \multicolumn{3}{c|}{\textbf{Overlap}} & \multicolumn{2}{c|}{\textbf{Unique}} \\ \cline{2-16}
\multicolumn{1}{|c|}{} & \multicolumn{1}{c|}{\textbf{GT}} & \multicolumn{1}{c}{\textbf{TP}} & \multicolumn{1}{c}{\textbf{FP}} & \multicolumn{1}{c}{\textbf{FN}} & \multicolumn{1}{c}{\textbf{P(\%)}} & \multicolumn{1}{c|}{\textbf{R(\%)}} & \multicolumn{1}{c}{\textbf{No.I}} & \multicolumn{1}{c}{\textbf{No.}} & \multicolumn{1}{c}{\textbf{TP}} & \multicolumn{1}{c|}{\textbf{P(\%)}} & \multicolumn{1}{c}{\textbf{+S}} & \multicolumn{1}{c}{\textbf{+D}} & \multicolumn{1}{c|}{\textbf{Eq}} & \multicolumn{1}{c}{\textbf{S}} & \multicolumn{1}{c|}{\textbf{D}} \\ \hline
\textbf{G.C} & \multicolumn{1}{r|}{100} & 94.6 & 7.4 & 4.4 & 92.7 & 95.6 & 193 & 97 & 97 & 100 & 6.8 & 4.0 & 49.4 & 34.4 & 36.8 \\
\textbf{G.G} & \multicolumn{1}{r|}{157} & 149 & 4.2 & 7.0 & 97.3 & 94.5 & 145 & 79 & 68 & 86.1 & 2.8 & 3.0 & 43.6 & 99.6 & 18.6 \\
\textbf{A.M} & \multicolumn{1}{r|}{59} & 43.8 & 4.4 & 9.2 & 90.9 & 77.9 & 137 & 63 & 31 & 49.2 & 8.8 & 2.6 & 3.6 & 28.8 & 16.0 \\
\textbf{M.C} & \multicolumn{1}{r|}{50} & 35.4 & 4 & 9.2 & 89.8 & 74.1 & 104 & 55 & 31 & 56.4 & 6.2 & 1.0 & 8.0 & 20.2 & 15.8 \\
\textbf{O.I} & \multicolumn{1}{r|}{15} & 11.8 & 0.8 & 2.4 & 93.7 & 78.7 & 17 & 14 & 12 & 85.7 & 3.2 & 1.0 & 2.2 & 5.4 & 5.6 \\
\textbf{O.S} & \multicolumn{1}{r|}{6} & 5.2 & 1.8 & 0.8 & 74.3 & 86.7 & 6 & 4 & 3 & 75.0 & 1.0 & 1.0 & 1.0 & 2.2 & 0.0 \\
\textbf{S.C} & \multicolumn{1}{r|}{27} & 19.8 & 6.4 & 5.4 & 75.6 & 74.4 & 41 & 21 & 21 & 100 & 6.6 & 1.0 & 2.0 & 10.2 & 11.4 \\
\textbf{S.T} & \multicolumn{1}{r|}{21} & 17.8 & 4.0 & 4.2 & 81.7 & 80.2 & 50 & 24 & 19 & 79.2 & 2.8 & 5.8 & 3.8 & 5.4 & 6.6 \\
\textbf{S.A} & \multicolumn{1}{r|}{16} & 15.4 & 3.4 & 1.0 & 81.9 & 93.9 & 68 & 24 & 19 & 79.2 & 0.0 & 3.8 & 8.2 & 3.4 & 7.0 \\
\textbf{Y.B} & \multicolumn{1}{r|}{4} & 3.2 & 1.8 & 0.0 & 64.0 & 80.0 & 60 & 14 & 8 & 57.1 & 0.0 & 0.0 & 1.0 & 2.2 & 7.0 \\
\textbf{Y.V} & \multicolumn{1}{r|}{161} & 134.2 & 8.4 & 23.0 & 94.1 & 84.5 & 132 & 78 & 49 & 62.8 & 7.2 & 0.0 & 10.4 & 116.6 & 31.4 \\ \hline
\textbf{Tot.} & \multicolumn{1}{r|}{\textbf{616}} & \textbf{530.2} & \textbf{46.6} & \textbf{65.8} & \textbf{85.1} & \textbf{83.7} & \textbf{953} & \textbf{473} & \textbf{358} & \textbf{75.7} & \textbf{45.4} & \textbf{23.2} & \textbf{133.2} & \textbf{328.4} & \textbf{156.2} \\ \hline
\end{tabular}
\begin{tablenotes}
    \footnotesize
    \item G.C: Github\_createOrgRepo, G.G: Github\_getOrgRepo, A.M: Ama\-deus\-Ho\-tel\_getMultiHotel, M.C: Marvel\_getComicIndividual, O.I: OMDB\_byIdOrTitle, O.S: OMDB\_bySearch, S.C: Spotify\_createPlaylist, S.T: Spotify\_getAlbumTracks, S.A: Spotify\_getArtistAlbums, Y.B: Yelp\_getBusinesses, Y.V: YouTube\_listVideos,  
    GT: \# constraints in the ground truth, (\# of {\tool} constraints grouped by variables=TP+FP). No.I: \#r of invariants. No: \# of invariants grouped by variables.
    +S: {\tool} better, +D: AGORA+ better, Eq: Equivalent. S: {\tool}, D: AGORA+.
\end{tablenotes}
\end{table}

%% file: tables/rq1_constraints_mining_ours_dataset.tex
\begin{table}[t]
\centering
\tabcolsep 0.8pt
\caption{Constraints Mining on the \underline{{\tool} dataset} (RQ1)}
\vspace{-9pt}
\small
\label{tab:rq1-constraint-mining-ours-dataset}
\footnotesize
\begin{tabular}{|l|rrrrrr|rrrr|rrr|rr|}
\hline
\multicolumn{1}{|c|}{\multirow{2}{*}{\textbf{API}}} & \multicolumn{6}{c|}{\textbf{{\tool}}} & \multicolumn{4}{c|}{\textbf{AGORA+}} & \multicolumn{3}{c|}{\textbf{Overlap}} & \multicolumn{2}{c|}{\textbf{Unique}} \\ \cline{2-16}
\multicolumn{1}{|c|}{} & \multicolumn{1}{c|}{\textbf{GT}} & \multicolumn{1}{c}{\textbf{TP}} & \multicolumn{1}{c}{\textbf{FP}} & \multicolumn{1}{c}{\textbf{FN}} & \multicolumn{1}{c}{\textbf{P(\%)}} & \multicolumn{1}{c|}{\textbf{R(\%)}} & \multicolumn{1}{c}{\textbf{No.I}} & \multicolumn{1}{c}{\textbf{No.}} & \multicolumn{1}{c}{\textbf{TP}} & \multicolumn{1}{c|}{\textbf{P(\%)}} & \multicolumn{1}{c}{\textbf{+S}} & \multicolumn{1}{c}{\textbf{+D}} & \multicolumn{1}{c|}{\textbf{Eq}} & \multicolumn{1}{c}{\textbf{S}} & \multicolumn{1}{c|}{\textbf{D}} \\ \hline
\textbf{C.H} & \multicolumn{1}{r|}{42} & 40.8 & 6.0 & 1.2 & 87.2 & 97.1 & 25 & 23 & 23 & 100 & \textbf{3.2} & 0.0 & 7.2 & \textbf{30.4} & 12.6 \\

\textbf{G.B} & \multicolumn{1}{r|}{61} & 46.2 & 3.2 & 14.8 & 93.5 & 75.7 & 215 & 123 & 91 & 74.0 & \textbf{7.8} & 6.0 & 17.8  & 14.6 & \textbf{59.4}   \\

\textbf{G.C} & \multicolumn{1}{r|}{86} & 67.0 & 3.2 & 19.4 & 95.4 & 77.5 & 313 & 178 & 128 & 71.9 & \textbf{11.2} & 3.2 & 25.4 & 27.2 & \textbf{88.2} \\

\textbf{G.G} & \multicolumn{1}{r|}{119} & 78.2 & 5.6 & 41.0 & 93.3 & 65.6 & 550 & 318 & 225 & 70.8 & \textbf{17.6} & 4.0 & 38.2 & 18.4 & \textbf{165.2} \\

\textbf{G.I} & \multicolumn{1}{r|}{268} & 156.6 & 3.2 & 111.8 & 98.0 & 58.3 & 1239 & 637 & 451 & 70.8 & \textbf{38.2} & 10.0 & 44.2 & 64.2 & \textbf{358.6} \\

\textbf{G.P} & \multicolumn{1}{r|}{248} & 177.6 & 4.6 & 71.2 & 97.5 & 71.4 & 1228 & 621 & 376 & 60.5 & \textbf{30.8} & 2.0 & 66.2 & 78.6 & \textbf{277.0} \\

\textbf{G.R} & \multicolumn{1}{r|}{55} & 44.0 & 3.4 & 11.0 & 92.8 & 80.0 & 220 & 122 & 88 & 72.1 & \textbf{7.2} & 2.0 & 19.2 & 15.6 & \textbf{59.6} \\

\textbf{S} & \multicolumn{1}{r|}{107} & 88.4 & 18.8 & 20.8 & 82.5 & 81.0 & 123 & 86 & 62 & 72.1 & \textbf{27.4} & 1.6 & 18.4 & \textbf{41.0} & 14.6 \\ \hline

\textbf{Tot.} & \multicolumn{1}{r|}{986} & 698.8 & 48.0 & \textbf{291.0} & {\bf 93.6} & 70.6 & 3913 &  2108 & 1444 & {\bf 74.0} & \textbf{143.4} & 28.8 & 236.6 & 290.0 & \textbf{1,035.2} \\ \hline
\end{tabular}

\begin{tablenotes}
    \footnotesize
    \item CH: Canada Holidays, GB: GitLab Branch, GC: GitLab Commit, GG: GitLab Groups, GI: GitLab Issues, GP: GitLab Project, GR: GitLab Repository, S: Stripe.GT: \# constraints in the ground truth,  (\# of {\tool} constraints grouped by variables=TP+FP). No.I: \#r of invariants. No: \# of invariants grouped by variables.
    +S: {\tool} better, +D: AGORA+ better, Eq: Equivalent. S: {\tool}, D: AGORA+.
\end{tablenotes}
\end{table}

%% file: sections/detected-constraint-types-2.tex
\input{tables/types-of-detected-constraints}

\subsubsection{Analysis on Types of Constraints Detected by {\tool}}

As shown in Table~\ref{tab:mining-constraint-by-category}, we summarize the distribution of 1,177 true positive constraints identified across 16 main categories in both the AGORA and {\tool} datasets. Among these, the I/O constraint type, which represents logical and dependency relationships between input and output fields within individual APIs, accounts for 309 instances (26.3\%). This indicates that {\tool} is effective at mining commonly described I/O relationships within APIs.



The isUrl category constitutes the second largest group, comprising 290 constraints (24.6\%), showing the widespread presence of URL-related attributes in the API specifications. The isDateTime constraints, which specify date/time formats and rules, rank next with 160 constraints (13.6\%). The Value-In-Set category, containing 150 constraints (12.7\%), reflects the common use of enumerated or bounded value restrictions that encode domain-specific semantics such as status codes, country abbreviations, or numeric thresholds in the APIs. Following these, the Composite category includes 87 constraints (7.4\%), showing that {\tool} can capture complex logical conditions composed of multiple atomic constraints.

The remaining categories account for about 15.4\% of all detected constraints, covering specialized rules such as date formats, string lengths, numeric ranges, and array checks, which reflect finer-grained or domain-specific properties. This result demonstrates that {\tool} can detect a wide variety of common oracles found in real-world API schemas and response bodies.

%% file: tables/types-of-detected-constraints.tex
\begin{table}[t]
\centering
\tabcolsep 1.5pt
\caption{Types of Constraints detected by {\tool}}
\vspace{-7pt}
\small
\label{tab:mining-constraint-by-category}
\footnotesize
\begin{tabular}{|c|l|r!{\vrule width 1.2pt}c|l|r|}
\hline
\textbf{No.} & \textbf{Category/Test Oracles} & \textbf{Count} &
\textbf{No.} & \textbf{Category/Test Oracles} & \textbf{Count} \\ \hline
1 & I/O & 309 & 9 & isBoolean & 18 \\ \hline
2 & isUrl & 290 & 10 & isNumber & 18 \\ \hline
3 & isDateTime & 160 & 11 & isUnixTime & 14 \\ \hline
4 & Value-In-Set & 150 & 12 & Template Literals & 11 \\ \hline
5 & Composite & 87 & 13 & ArrayTypeOracle\_isString & 3 \\ \hline
6 & isDate & 45 & 14 & Array\_Specific\_Sizes & 3 \\ \hline
7 & String\_Specific\_Length & 35 & 15 & N-ary atomic & 3 \\ \hline
8 & Value-In-Range & 29 & 16 & isTime & 2 \\ \hline
\end{tabular}
\end{table}


%% file: sections/complete-eval-new.tex
\subsection{Constraint Test Generation Accuracy (RQ2)}
\label{sec:rq2}


\subsubsection{Methodology} 
{\color{black}{Using the best-performing result from Section~\ref{sec:rq1}, we generated validation scripts for all the mined constraints by following the process in Figure~\ref{fig:test-gen}.}} We manually evaluated whether each verification script correctly reflects the intended logic described by its corresponding constraint. The evaluation followed the criteria mentioned during the construction of the ground truth as in Section~\ref{sec:eval}. Verification scripts were executed against real-world API responses. Each constraint was verified by 50 API responses.







\subsubsection{Results}For the AGORA/AGORA+ dataset, as seen in Table~\ref{tab:rq2-3-4-agora}, {\tool} performs well in generating test scripts from mined constraints, achieving a validity rate of 86.4\% and covering 76.6\% of the ground-truth constraints. However, it missed 135 out of 578 constraints. Notably, for some services, e.g., Github\_createOrgRepo and Github\_getOrgRepo, {\tool} achieved up to 82\% recall, while for Yelp, it reached 100\% recall.

For the {\tool} dataset, Table~\ref{tab:rq2-3-4-rbc} shows a precision of 91.7\%, indicating that most generated test scripts correctly validate their associated constraints. The recall reached 71.3\%, though 279 out of 973 constraints were missed. These metrics are slightly lower than those in RQ1, as this evaluation jointly considers both constraint correctness from the mining and the correctness of test execution.

\tool{} excels in test generation for {\em Response Property constraints}, which often involve clear format or range-based rules, e.g., ``three-letter currency code'' or ``Unix epoch timestamp.'' These constraints are easier to detect due to their descriptive nature. In contrast, {\em Request--Response}, {\em Range-of-Value}, and {\em Computed Value} constraints are more error-prone, as they require accurate mappings between request parameters and response properties. However, \tool{} still achieves very high precision on these constraints.

\input{tables/rq2-3-4-agora}

\input{tables/rq2-3-4-our}

Test generation errors often stem from missing or vague descriptions, especially in services like GitLab, where response property documentation is sparse. For example, the request parameter \code{avatar} (used for image uploads) was incorrectly mapped to the response field \code{avatar\_url} due to name similarity, despite different semantics.
Services that define complex {\em Data Format} constraints---such as ``\code{numberOfNights}  must always be greater than 0,'' or fields expected to be unique like \code{id}, \code{code} or \code{barCode}---also present challenges when descriptions are missing or implicit. Attributes with special rules ({\em Name-based constraint}) like \code{latitude} or \code{longitude} in \code{AmadeusHotel\_get\-MultiHotel} or \code{lang} in Marvel were either overlooked or led to test scripts that failed to express the intended constraint logic. In these cases, the lack of structured or informative descriptions in the specification hindered constraint detection and test generation.

%% file: tables/rq2-3-4-agora.tex
\begin{table}[t]
\tabcolsep 2.5pt
\footnotesize
\caption{Constraint Test Generation (RQ2), Test Generation from Correct Constraints (RQ3), and Test Outcomes (RQ4) on the \underline{AGORA/AGORA+} Dataset.}
\label{tab:rq2-3-4-agora}
\vspace{-9pt}
\begin{tabular}{|l|r|r|r|r|r|r|r|r|c|r|r|r|}
\hline
\multirow{2}{*}{\textbf{API}} & \multicolumn{6}{c|}{\bf Constraint Test Gen (RQ2)} & \multicolumn{3}{p{2.5cm}|}{
\centering \textbf{Test Gen (RQ3)} 
\textbf{(Correct Constraints)}} & \multicolumn{3}{p{1cm}|} {\centering \textbf{Outcome} \textbf{(RQ4)}} \\
\cline{2-13}
& \textbf{TP} & \textbf{FP} & \textbf{FN} & \textbf{P\%} & \textbf{R\%} & \textbf{F1\%} & \textbf{N} & \textbf{\checkmark} & \textbf{P\%} & \textbf{\checkmark} & \textbf{$\times$} & \textbf{?} \\
\hline
\textbf{G.C}      & 78  & 12 & 17 & 86.7 & 82.1 & 84.3 & 79  & 78  & 98.7 & 77  & 0 & 1 \\
\textbf{G.G}     & 127 & 8  & 27 & 94.1 & 82.5 & 87.9 & 127 & 127 & 100.0 & 127 & 0 & 0 \\
\textbf{A.M}     & 23  & 15 & 22 & 60.5 & 51.1 & 55.4 & 29  & 23  & 79.3 & 16  & 2 & 5 \\
\textbf{M.C}      & 21  & 9  & 18 & 70.0 & 53.8 & 60.9 & 26  & 21  & 80.8 & 13  & 0 & 8 \\
\textbf{O.I}        & 12  & 0  & 3  & 100.0 & 80.0 & 88.9 & 12  & 12  & 100.0 & 8   & 4 & 0 \\
\textbf{O.S}       & 5   & 2  & 1  & 71.4 & 83.3 & 76.9 & 6   & 5   & 83.3 & 5   & 0 & 0 \\
\textbf{S.C}     & 17  & 7  & 8  & 70.8 & 68.0 & 69.4 & 19  & 17  & 89.5 & 15  & 0 & 2 \\
\textbf{S.T}    & 19  & 0  & 6  & 100.0 & 76.0 & 86.4 & 19  & 19  & 100.0 & 10  & 0 & 9 \\
\textbf{S.A}   & 15  & 3  & 2  & 83.3 & 88.2 & 85.7 & 16  & 15  & 93.8 & 8   & 1 & 6 \\
\textbf{Y.B}        & 4   & 1  & 0  & 80.0 & 100.0 & 88.9 & 4   & 4   & 100.0 & 3   & 1 & 0 \\
\textbf{Y.V}     & 122 & 13 & 31 & 90.4 & 79.7 & 84.7 & 124 & 122 & 98.4 & 79  & 1 & 42 \\
\hline
\textbf{Total}       & 443 & 70 & 135 & \textbf{86.4} & \textbf{76.6} & \textbf{81.2} & 461 & 443 & \textbf{96.1} & 361 & \textbf{9} & 73 \\
\hline
\end{tabular}

\begin{tablenotes}
    \small
    \item For test outcomes: \textbf{\checkmark} (Matched), \textbf{$\times$} (Mismatched), and \textbf{?} (Unknown).
\end{tablenotes}
\end{table}

%% file: tables/rq2-3-4-our.tex
\begin{table}[]
\tabcolsep 2.5pt
\footnotesize
\caption{Constraint Test Generation (RQ2), Test Generation from Correct Constraints (RQ3), Test Outcomes (RQ4) on the \underline{{\tool}} Dataset.}
\label{tab:rq2-3-4-rbc}
\vspace{-9pt}
\begin{tabular}{|l|r|r|r|r|r|r|r|r|c|r|r|r|}
\hline
\multirow{2}{*}{\textbf{API}} & \multicolumn{6}{c|}{\bf Constraint Test Gen (RQ2)} & \multicolumn{3}{p{2.5cm}|}{
\centering \textbf{Test Gen (RQ3)}

\textbf{(Correct Constraints)}
}
 & \multicolumn{3}{p{1cm}|}{\centering \textbf{Outcome} \textbf{(RQ4)}} \\
\cline{2-13}
& \textbf{TP} & \textbf{FP} & \textbf{FN} & \textbf{P\%} & \textbf{R\%} & \textbf{F1\%} & \textbf{N} & \textbf{\checkmark} & \textbf{P\%} & \textbf{\checkmark} & \textbf{$\times$} & \textbf{?} \\
\hline
\textbf{C.H} & 36  & 0  & 6   & 100 & 85.7 & 92.3 & 36  & 36  & 100.0 & 22  & 2  & 12 \\
\textbf{G.B}  & 47  & 6  & 13  & 88.2  & 75.0 & 84.7 & 47  & 45  & 95.7  & 42  & 1  & 2  \\
\textbf{G.C}  & 64  & 7  & 16  & 90.1  & 75.3 & 88.5 & 69  & 64  & 92.8  & 51  & 6  & 7  \\
\textbf{G.G}  & 89  & 6  & 26  & 93.7  & 76.1 & 85.8 & 91  & 89  & 97.8  & 78  & 6  & 4  \\
\textbf{G.I}  & 171 & 8  & 91  & 95.5  & 64.3 & 78.7 & 175 & 171 & 97.7  & 132 & 19 & 20 \\
\textbf{G.P} & 163 & 16 & 70  & 91.1  & 68.5 & 80.6 & 168 & 163 & 97.0  & 139 & 0  & 24 \\
\textbf{G.R}   & 44  & 7  & 11  & 86.3  & 80.0 & 83.0 & 46  & 44  & 95.7  & 41  & 0  & 2  \\
\textbf{S}    & 82  & 13 & 26  & 86.3  & 74.5 & 82.0 & 84  & 82  & 97.6  & 68  & 3  & 11 \\
\hline
\textbf{Total}     & 694 & 63 & 259 & {\bf 91.7}  & {\bf 71.3} & {\bf 82.5} & 716 & 694 & {\bf 96.9}  & 573 & 37 & 82\\
\hline
\end{tabular}

\begin{tablenotes}
    \small
    \item For test outcomes: \textbf{\checkmark} (Matched), \textbf{$\times$} (Mismatched), and \textbf{?} (Unknown).
\end{tablenotes}
\end{table}

%% file: sections/test-gen-only-eval.tex
\subsection{Test Generation Accuracy from Correctly Mined Constraints (RQ3)}

\subsubsection{Methodology}
The experiment in RQ2 evaluates the entire process from constraint mining to test generation.
This RQ3 focuses only on evaluating the test generation for the constraints {\em correctly mined by {\tool}}. We perform test generation from those correct constraints. The generated test cases are manually evaluated if they correctly verify the associated constraints.
We used the following rules to check if a test case is correct:

\textbf{(1) Test Input:} The generated test case is correct if it correctly receives two inputs for Request-Response constraints: 1) requested information and 2) response data, and one input for Response Property constraints (response data).
    
\textbf{(2) Constraint Handling:} The generated test case must cover all conditions in the constraint.
    
\textbf{(3) Test Output:} Each script must be evaluated on 50 example responses collected from real-world data. The test must return:
    
\indent       i) 0 (\code{unknown}) if lacking of sufficient data for condition checking (e.g., empty or \code{null} value). 
        
        ii) 1 (\code{matched}) if the provided input satisfies the constraint. 
        
        iii) -1 (\code{mismatched}) if the provided input does not satisfy it.  
    


The final result is considered \code{matched} only if all test cases return 1 and none return -1; any case with 0 is acceptable. If all cases return 0, the result is \code{unknown}. If any case returns -1, the result is considered as \code{mismatched}. We only consider the set of test cases derived from valid constraints as identified in RQ2 (denoted as \(N\) in Tables~\ref{tab:rq2-3-4-agora}--\ref{tab:rq2-3-4-rbc}). 


\vspace{-4.5pt}
\subsubsection{Results}

As seen in Table~\ref{tab:rq2-3-4-rbc}, {\tool} generates 714 TP test cases out of 746 in the {\tool} dataset, resulting a precision of 94.3\%. In 4/8 services, it achieves a precision of 96\%, while the lowest precision is in \code{GitLab Repository API}, with a precision of 83\%. For the AGORA/AGORA+ dataset, it generates 443 correct test cases out of 513, with a precision of 96.1\%. In two services, it achieves~100\%. 

Note that in RQ2, we consider all mined constraints, while in RQ3, we use only the correctly mined ones. Thus, the results in RQ3 (e.g., precision of 96.1\% on the AGORA dataset) are higher than those in RQ2 (e.g., precision of 86.4\%).


Our analysis reveals that \tool excels in generating code for {\em Response Property constraints}, which mainly involve format or value range validation. In contrast, {\em Request-Response constraints} are more error-prone due to the complex logic required to parse and verify dependencies between request parameters and response properties. Further analysis reveals that test generation errors primarily stem from (1) missing descriptions and (2) ambiguous keywords. 



Consider the \code{`get-/issues'} endpoint in GitLab Issues, where a constraint exists between the request parameter \code{`due\_date'} and a response property of the same name. The parameter is described as \textit{``Accepts: 0 (no due date), overdue, week, month, next\_month,''} while the response property lacks a description. As a result, our tool generates test cases to validate the property against this constraint, even though \code{`due\_date'} in the response is a date-time string, leading to incorrect tests. Such errors are common in GitLab due to insufficient descriptions for response properties.

In the same \code{`get-/issues'} endpoint, the \code{`milestone'} parameter is described as \textit{``The milestone title. None lists all issues with no milestone. Any lists all issues that have an assigned milestone.''} The API filters returned issues based on \code{`None'} or \code{`Any'} from the request parameter. However, {\tool} misinterpreted \code{`None'} as Python's \code{null}, leading to errors. This mistake in \code{`milestone'} propagated to 13 other incorrect test cases due to its involvement in multiple~operations.

%% file: sections/response-body-testing.tex
\subsection{Usefulness: Detecting Mismatches between Constraints and API Responses (RQ4)}


\subsubsection{Methodology}
{\color{black}{
In this experiment, we use {\tool}'s generated test cases to detect mismatches between the API specification and API response bodies. A mismatch often indicates either that 1) the specification is outdated; or 2) the APIs' implementation does not align with the constraints in their specification.
To this goal, we performed the following. For the {\tool} dataset, 
for each API endpoint, we used RESTest~\cite{MartinLopez2021Restest-2} to generate 50 API requests, executed the SUT with APIs, and obtained the actual responses. For each execution, we collected 1) request data, and 2) actual response data. We used only the pairs of request-response with the status code of 2xx. Response data contains multiple properties, each attached with constraints and their test cases generated by {\tool}. We ran the test cases (in RQ3) {\em generated for the constraints on the properties in the response data to collect the outcomes}. If the result is false, we report a mismatch. Otherwise, it is a match.

For the AGORA/AGORA+ dataset, we used the request-response pairs provided on its repository, whose responses were obtained by its authors from the actual execution. We ran the test cases generated by {\tool} and compared with the responses as described.

}}

\subsubsection{Results}

For the {\tool} dataset, we performed test runs on all 694 correctly generated test cases from RQ3  (Table~\ref{tab:rq2-3-4-rbc}). The tests evaluate if the response data conforms to the constraints detected from the API specification. 
Our test results indicate 573 ‘matched’ response bodies (i.e., consistent with the specification), 37 ‘mismatched’, and 82 ‘unknown’. Out of 714 correctly mined constraints, 573 were verified by the test cases, meaning that 82.5\% of the constraints are met by the actual execution of the SUTs. Our tool detected {\em 37 mismatches}, revealing inconsistencies between specifications and execution of the APIs. An ‘unknown’ occurs when a property is absent in the response body due to its optional nature.

For the AGORA/AGORA+ dataset, we performed test runs on all 443 correctly generated tests from RQ3  (Table~\ref{tab:rq2-3-4-agora}). Our test results indicate 361 ‘matched’ response bodies (i.e., consistent with the specification), 9 ‘mismatched’, and 73 ‘unknown’. Out of 443 correctly mined constraints, 361 were verified by the test cases, meaning that 81.5\% of the constraints are satisfied by the actual execution of the APIs. Our tool detected {\em 9 mismatches}, revealing inconsistencies between specifications and API~executions.


\input{sections/detected-mismatch-types-2}

\subsubsection{Root-cause Analysis} We reported the root causes of these mismatches: (1) incompatible~data formats, (2) implicitly-described nullable properties, and (3) inter-parameter request dependencies. 



\vspace{1pt}
\textbf{(1) Incompatible Data Formats}: Our tool detected constraint mismatches mainly from GitLab services. For instance, the constraint {\em "date will be returned in ISO 8601 format \code{YYYY-MM-DDTHH:MM"}} appears in GitLab operations. However, the actual data format is \code{"2012-09-20T08:50:22.000Z"}, which includes a decimal part for seconds, leading to an inconsistency. 
Interestingly, we found that three instances of this type of inconsistency {\bf were reported as the issues} on the GitLab forum~\cite{gitlab_issue_time1, gitlab_issue_time2, gitlab_issue_time3}. This is {\em anecdotal evidence on {\tool}'s usefulness in detecting real-world data-format issues}. 


\vspace{4pt}
\textbf{(2) Implicitly-Described, Nullable Properties}: This issue is common in GitLab, where some response properties are nullable but lack descriptions in the specification, leading to parsing errors. Notably, these issues have been reported on the GitLab forum~\cite{gitlab}.

\vspace{4pt}
\textbf{(3) Inter-Parameter Request Dependencies}: This was found in only one case. For the operation \code{'GET/groups'} from GitLab Group, there is a constraint on the request parameter \code{"order\_by"} affecting the \code{"name"} property in the response data. This logic dictates that the array of groups in the API response should be sorted according to the \code{"order\_by"} parameter. {\tool} checked only one-to-one conditions among parameters, whereas the sorting order depends on both \code{"order\_by"} and \code{"sort"}~parameters (specifying the sort direction). As a result, this resulted in a detected mismatch. Future work could address more complex dependencies among multiple parameters.


%% file: sections/detected-mismatch-types-2.tex
\subsubsection{Analysis on Detected Mismatches}

{\color{custom-blue}{
For 46 mismatches detected by {\tool}~\cite{detected-bugs}, we conducted a detailed analysis by classifying them into 6 constraint categories, as illustrated in Figure ~\ref{fig:bugclassify}. The isDateTime category, which refers to the constraint in date and time data types, accounts for the largest proportion (45.7\%), indicating that temporal inconsistencies—such as variations in date–time formats or missing temporal fields—are the predominant mismatch between API specifications and their actual responses. 

\begin{wrapfigure}{r}{2.5in}
  \centering
  \includegraphics[width=2.5in]{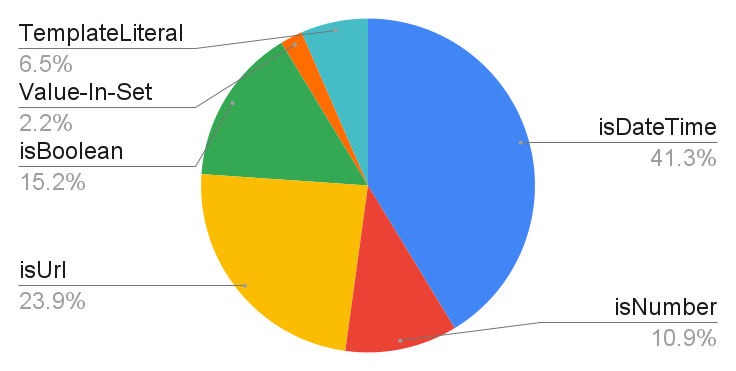}
 \vspace{-18pt}
  \caption{Detected Mismatches by Category}
  \label{fig:bugclassify}
\end{wrapfigure}

Other categories, including IsUrl (the property as a URL) and IsNumber (each 13.0\%), IsBoolean and boolean values (10.9\%), and Template Literals (e.g., String templates/lengths:  Canadian provincial abbreviations in Canadian\_Holidays API), and Value-In-Set (8.7\%, e.g., in YouTube, video length in ISO\_8601 duration formats: PT\#M\#S), show  lower but still notable~occurrences. 

These findings suggest that the detected mismatches are non-trivial, and they mainly arise in the fields governed by strict formatting or parsing rules. In particular, the prevalence of date–time errors highlights the need for better normalization and schema-level validation. Beyond format-related issues, several mismatches may also stem from incorrect implementation or schema drift, where API implementations change faster than their specifications. This result indicates that {\em {\tool} is useful in continuous validation and automated schema synchronization and in maintaining consistency between API behavior and its declared constraints.}

}}

%% file: sections/ablation.tex
\subsection{Ablation Study (RQ5)}






In this experiment, we aim to compare two variants: 1) ({\tool}$^{-}$): {\tool} without observation-confirmation (OC) prompting and semantic verifier, and 2) {\tool} with both components. To build the former, we modify the prompt for constraint mining in Figure~\ref{prompt:MAPPING_PROMPT} as follows. Instead of providing GPT-4o with observations derived from another prompt, we feed the description of the parameters and response schema as in the API specification. This prompt replaces Blocks \circlednum[request]{1}-\circlednum[request]{5} from Figure~\ref{fig:method-overview}. After providing data on a property, we instruct the LLM to decide if the given property contains a constraint and if there is enough to verify it. This modification merges steps \circlednum[response]{6} and \circlednum[response]{7} into a single step. The output is \textit{yes} or \textit{no}. We ran both variants five times and recorded the mean results in the same process as described in the previous sections.

\begin{table}[t]
\small
\centering
\caption{Ablation Study on the \underline{{\tool} dataset} (RQ5)}
\vspace{-5pt}
\label{tab:ablation}
\begin{tabular}{|c|rrrrrr|}
\hline
\multirow{2}{*}{\textbf{Variant}}
  & \multicolumn{6}{c|}{\textbf{Constraint Test Generation}} \\ \cline{2-7}
  & \textbf{TP} & \textbf{FP} & \multicolumn{1}{c|}{\textbf{FN}}
  & \textbf{P} & \textbf{R} & \textbf{F1} \\ \hline
\textbf{{\tool}-} & 846  & 279 & \multicolumn{1}{r|}{706} & 75.2\% & 54.5\% & 63.2\% \\ \hline
\textbf{{\tool}} & 1,137 & 133 & \multicolumn{1}{r|}{394} & 89.5\% & 74.3\% & 81.2\% \\ \hline
\end{tabular}
\end{table}

As seen in Table~\ref{tab:ablation}, the elimination of observation-confirmation prompting and semantic verifier affects the outcomes, as evidenced by a reduction of 312 correct constraints. Concurrently, there is an increase in the number of false positives. The F1-score and recall of {\tool}- decreased by approximately 25\% compared to the original {\tool}. False positives frequently arise when the model mis-maps request parameters to response properties based solely on their names. In addition, there are cases where certain fields have names that misleadingly suggest they represent constraints, but based on their descriptions and the predefined evaluation rules, they are actually not constraints. Such mistakes are less common in {\tool}, where the observation prompt is used to enhance the property before it is processed by the confirmation prompt. {\em This result confirms our key contribution of our LLM prompting strategy}.

The semantic verifier's goal is to remove the invalid constraints to improve precision. We used a simple verifying mechanism via examples in API specification (Section~\ref{sec:ideas}). We removed the semantic verifier and ran the static component. In the \tool dataset with 89 examples, one example invalidated one detected constraint. In the AGORA dataset, with 223 examples over 11 API operations, 6 examples invalidated 6 false-positive constraints. Overall, the verifier accurately confirmed all valid constraints and successfully eliminated 7 incorrect constraints, thus achieving higher precision (86.4\% increasing to 87.5\%) while maintaining recall 76.6\%. These results show that more examples in the API are useful to invalidate more incorrect constraints and confirm the correct ones. In general, other types of semantic verifier can be integrated into our framework such as constraint solvers, SMT solvers, domain-specific checkers (valid zip code, phone number format, or valid date checkers), etc.

%% file: sections/cost-analysis-2.tex
\subsection{Token Cost Analysis}

For {\bf constraint inference}, {\tool} does not rely on LLMs to analyze API responses; it instead operates directly on the API specification. This allows each API response field to be processed only once to generate test oracles, which can then be reused across all subsequent API calls. We measured the marginal inference cost of \tool with GPT-4o. Due to that strategy, for 19 APIs from both datasets, {\tool} consumed 1,300,479 input tokens (68,445 per API) and 180,527 output tokens (9,501 per API). The total cost for both datasets was~\$5.1. \footnote{\$2.5 per 1M input tokens (\$1.25 if cached) \$10 per 1M output tokens.} 

The {\bf test generation} phase is more resource-intensive, consuming a total of 1,332,494 input tokens and 473,521 output tokens, with the former accounting for around 73\% of the total. The total LLM cost was estimated at \$8.1. The highest token consumption occurred in the GitLab APIs and YouTube\_listVideos, with the GitLab Issues API alone using 199,361 input tokens and 70,981 output tokens. Smaller APIs such as Yelp\_getBusinesses and OMDB required fewer numbers of tokens.


Overall, the analysis of token usage and monetary cost shows {\tool}'s efficiency, with 
the cost of \$13.2 in two phases of constraint inference and test script generation for all APIs in the~datasets. 


%% file: sections/threats-to-validity.tex
\section{Threats to Validity}
{\em 1. Internal Validity.} LLM hallucinations might lead to unexpected outcomes~\cite{sallou2024breaking}. To mitigate this, we incorporated (1) temperature of zero for GPT-4o, (2) a semantic verifier (Figure~\ref{fig:test-gen}) and (3) observation-confirmation prompting (Figure~\ref{fig:method-overview}), which enhance validation through external API specifications and internal consistency checks. 

Data leakage is expected to be minimal: while specifications and APIs may appear in pre-training data, the specific pairs of specifications and constraints are not. To address non-determinism from input variation, we consistently used the same prompt templates.

{\em 2. External Validity.} Our dataset may not fully represent the API landscape, affecting generalizability. Outcomes could vary with different datasets, particularly those with unique constraints. Generated test cases may miss general constraints or edge cases beyond what is explicitly documented. Implicit constraints not in API specifications might also be valid but unaccounted for. External tools may introduce inaccuracies, potentially affecting the results.

%% file: sections/discussion.tex
\section{Discussion}
\label{sec:discussion}


Our goal and empirical results indicate that the static approaches such as {\tool} and the dynamic approaches are {\em complementary to each other}.
First, they complement each other in {\em different usage scenarios}. The API specifications would be available when 1) API developers provide the OAS specification files for API usages, 2) a tester performs testing based on OAS files, or 3) the front-end developers need to know the specification for back-end integration. If an OAS file is not available, a dynamic approach could extract the constraints from runtime information, e.g., in the regression testing or other scenarios where the application could be correctly executed. In contrast, if the application is under development and non-executable, {\tool} could rely on the OAS files.

Second, they are also {\em complementary by exploring different information channels}. {\tool} detects constraints from API specifications while a dynamic approach detects invariants through API execution. 
They provide coverage where the other falls short, either when specifications are incomplete or when execution history is overly specific or unreliable (e.g., during development). 

Third, as demonstrated in our {\em results}, each approach recovers shared constraints as well as unique ones: {\tool} covers cases missed by runtime test cases or when running is infeasible, while AGORA/AGORA+ handles cases absent from the specification. 

Finally, {\em for the mismatch/bug detection application}, they are also complementary. {\tool} detects bugs/mismatches between API specification and implementation. A dynamic one detects bugs across unseen responses in different executions.~Thus, we call for a combination between {\tool} and dynamic approaches~\cite{huynh2025rbctestleveragingllmsverify}.


%% file: sections/conclusion.tex
\section{Conclusion}

{\bf Novelty}. This work, {\tool}, demonstrates the effectiveness of leveraging LLMs to statically mine logical constraints to be used as oracles for API response bodies directly from API specifications, addressing key limitations of dynamic analysis-based approaches. By introducing the Observation-Confirmation prompting scheme, we reduce hallucinations and significantly improve the precision of mined oracles. {\tool} applies LLMs in combination with API documentation to derive {\em the oracles from the response bodies} without relying on runtime execution, enabling broader and more accurate constraint discovery. This novel use of LLMs for static oracle mining represents a shift from traditional behavior-driven testing toward specification-driven validation. {\tool} uncovers {\em 46 real-world inconsistencies} between documented specifications and actual APIs. 



\vspace{1pt}
{\bf Practical Impacts}. 
Our results also encourage further exploration into {\bf combining static and dynamic techniques}, leveraging the strengths of each, to build hybrid systems that can generate richer test oracles, increase coverage, and surface deeper specification-implementation mismatches across interfaces.